\pdfoutput=1

\documentclass[a4paper,11pt]{article}
\usepackage{jcappub} 
\usepackage{aas_macros}

\newcommand{\gr}{$\gamma$-ray}

\title{On detecting oscillations of gamma rays into axion-like particles  in turbulent and coherent magnetic fields} 

\author[a]{Manuel Meyer}
\author[b]{Daniele Montanino}
\author[a]{Jan Conrad}

\affiliation[a]{The Oskar Klein Center for CosmoParticle Physics, Department of Physics, Stockholm University, Albanova, SE-10691 Stockholm, Sweden}
\emailAdd{manuel.meyer@fysik.su.se}
\emailAdd{conrad@fysik.su.se}
\affiliation[b]{Dipartimento di Matematica e Fisica ``Ennio De Giorgi'', Universit\`{a} del Salento and Sezione INFN di Lecce, Via Arnesano, I-73100 Lecce, Italy}
\emailAdd{daniele.montanino@le.infn.it}

\arxivnumber{1406.5972}

\abstract{
Background radiation fields pervade the Universe, 
and above a certain energy any {\gr} flux emitted by an extragalactic source should be attenuated due to $e^+e^-$ pair production.
The opacity could be alleviated if photons oscillated into hypothetical axion-like particles (ALPs)
in ambient magnetic fields, leading to a {\gr} excess especially at high optical depths that could be detected
with imaging air Cherenkov telescopes (IACTs). 

Here, we introduce a method to search for such a signal in {\gr} data and to estimate sensitivities for future observations.
Different magnetic fields close to the {\gr} source are taken into account in which photons can convert into ALPs
that then propagate unimpeded over cosmological distances until they re-convert in the magnetic field 
of the Milky Way.
Specifically, we consider the coherent field at parsec scales in a blazar jet as well as the turbulent field 
inside a galaxy cluster.
For the latter, we explicitly derive the transversal components of a magnetic field with gaussian turbulence
 which are responsible for the photon-ALP mixing. 
To illustrate the method, we apply it to a mock IACT array with characteristics similar to the Cherenkov Telescope Array and investigate the
 dependence of the sensitivity to detect a {\gr} excess on the magnetic-field parameters.
}

\begin{document}
\keywords{gamma ray experiments, active galactic nuclei, magnetic fields, galaxy clusters}
\maketitle

\flushbottom

\section{Introduction} 
\label{sec:intro}
The observation of very high energy {\gr} emission (VHE; energy $E \gtrsim 100$\,GeV) from extragalactic  
sources is a unique probe of VHE photon propagation over cosmological distances.  
During the propagation to Earth, $\gamma$ rays can undergo pair production, $\gamma\gamma \to e^+e^-$,
with photons of the extragalactic background light (EBL) \cite{nikishov1962,jelley1966,gould1967}. 
This attenuation of the source-intrinsic flux scales exponentially with the optical depth $\tau$,
a monotonically increasing function with {\gr} energy, the source redshift, and the EBL photon density \cite[see][for a review]{dwek2013}.
The EBL ranges from ultraviolet (UV) to far infrared (IR) wavelengths
and comprises the starlight emitted by stars and starlight absorbed and re-emitted by dust in galaxies,
integrated over the entire history of the universe \cite[see][for a review]{hauser2001}.
Direct measurements of the EBL are extremely difficult due to the strong contamination with foreground emission \cite{hauser1998} but lower limits can be derived from galaxy number counts in the optical and infrared \cite[e.g.][]{madau2000,fazio2004}.

This standard scenario was questioned \cite{protheroe2000} after the HEGRA observations
of the blazar Markarian 501 above 20\,TeV \cite{mkn501hegra1999}.
The EBL models available at that time \cite[e.g.][]{malkan1998} predicted a high EBL photon density at IR wavelengths, rendering 
an observation of multi TeV $\gamma$ rays highly unlikely. 
Recent EBL models \cite[e.g.][]{kneiske2010,dominguez2011,gilmore2012,inoue2012}, however, more or less agree on a level of the background photon density close to the lower limits available from galaxy number counts,
reconciling the measurement with expectations. 
Nevertheless, the number of observations of blazars (active galactic nuclei, AGN, with their jet closely aligned along the line of sight) in the optical thick regime ($\tau > 2$) is steadily increasing \cite[e.g.][]{1es1101h2356hess2006,3c279magic2008,pg1553hess2008,pks1424veritas2010}, 
reaching optical depths $\tau \gtrsim 5$ \cite{furniss2013}.
A statistical analysis of more than 50 AGN VHE spectra showed a $4\,\sigma$ indication for an overestimation of the EBL attenuation 
 in the $\tau > 2$ regime (the significance is reduced to $2\,\sigma$ if systematic uncertainties are taken into account) \cite{horns2012}.
Furthermore, several sources show harder spectral indices than expected from EBL absorption \cite{deangelis2009,deangelis2011,deangelis2013}.
The level of the EBL photon density inferred from {\gr} absorption features in \textit{Fermi}-LAT \cite{ackermann2012ebl} and H.E.S.S. spectra \cite{hess2013ebl} is compatible with current EBL models. It should be noted that these analyses are dominated by data in the 
optical thin regime and that the allowed level is close to or slightly below the lower limits on the EBL 
 at intermediate redshifts, $0.2 \lesssim z \lesssim 0.5$ \cite{meyer2013pks1424}.

One possibility to account for a decreased opacity of VHE $\gamma$ rays is the oscillation of photons into 
hypothetical pseudo-Nambu-Goldstone bosons in the presence of ambient magnetic fields \cite{deangelis2007,deangelis2009,deangelis2011,sanchezconde2009,dominguez2011alps,meyer2013}.
Such particles arise if additional gauge symmetries to the Standard Model are broken \cite[see][for a review]{jaeckel2010}
and one prominent example of such a particle is the axion 
that solves the strong CP problem in QCD \cite{pq1977,weinberg1978,wilczek1978}.
For the axion, the coupling to photons $g_{a\gamma}$ 
and its mass $m_a$ are related through the scale $f_a$ at which the additional symmetry is broken. 
For general theories of pseudo-Nambu-Goldstone Bosons or axion-like particles (ALPs) this is not the case.
Axion-like particles, while not able to cure the strong CP problem, naturally arise in string theories \cite{witten1984,cicoli2012,ringwald2014}.
The coupling of axions or ALPs to photons is described by the Lagrangian,
\begin{equation}
\mathcal{L}_{a\gamma} = -\frac 1 4 g_{a\gamma} F_{\mu\nu}\tilde F^{\mu\nu}a,
\label{eqn:lagr-alps}
\end{equation}
where $F_{\mu\nu}$ is the electromagnetic field tensor, $\tilde{F}_{\mu\nu}$ its dual, and $a$ is the ALP field strength.
The non-observation of ALPs potentially created in the sun by the CAST experiment yields an upper bound 
of $g_{a\gamma} < 8.8\times10^{-11}\,\mathrm{GeV}^{-1}$ for $m_a \lesssim 0.02$\,eV at 95\,\% confidence \cite{cast2007}.
Interestingly, with the tentative BICEP-2 results on the tensor-to-scalar ratio \cite{bicep2},
scenarios are preferred in which the additional symmetry is broken after inflation. 
This leads to a lower bound on the coupling strength,
$g_{a\gamma} \gtrsim 10^{-14}\,\mathrm{GeV}^{-1}$,
for $g_{a\gamma} \sim \alpha / (2\pi f_a)$ with the fine structure constant $\alpha$,
if one additionally requires that ALP dark matter should not exceed the cold dark matter abundance.
 \cite[e.g.][]{visinelli2014,dias2014}.

If a $\gamma$ ray converts into an ALP, it will evade pair production. While the {\gr} flux is attenuated in the EBL interactions,
ALPs reconverting into photons in the vicinity of the Earth can give rise to a boost in the observed flux,
 especially in the optical thick regime.
The resulting spectral hardening is not predicted in simple emission scenarios of blazars, 
for instance in self-synchrotron-Compton models \cite[e.g.][]{bloom1996}
that often suffice to describe their broadband spectral energy distribution\footnote{
In more specific scenarios, a spectral hardening can be expected due to, e.g., internal absorption \cite{aharonian2008},
emission from multiple sites with narrow electron distributions \cite{lefa2011a}, or comptonization by an ultra-relativistic particle wind \cite{aharonian2002}.
}.
Simultaneous observations of the spectrum at low and high $\tau$ values 
is mandatory in order to quantify the potential level of discrepancy between EBL model predictions and measurements. 
In this regard, the planned Cherenkov Telescope Array (CTA) \cite{cta2011} with its low energy threshold of $\sim 50$\,GeV
will be perfectly suited for such observations (see ref. \cite{mazin2013} for possible EBL studies with CTA). 
The point-source sensitivity of CTA is expected to be a factor of 10 better than currently operating imaging air Cherenkov telescopes (IACTs)
\cite{cta2011}. 

Here, we consider different coherent and turbulent magnetic field scenarios  
and we explicitly derive formulas for the transversal components of magnetic fields with gaussian turbulence.
A likelihood ratio test is introduced to determine the sensitivity of a CTA-like array to detect 
a spectral hardening induced by the oscillations of $\gamma$ rays into ALPs.
The likelihood ratio test and the magnetic field scenarios can easily be used in the future  
to calculate the sensitivity of, e.g., H.E.S.S. II \cite{becherini2012} and CTA to detect a boost in {\gr} fluxes.

We begin the discussion with a short review of photon-ALP oscillations 
in Section \ref{sec:alps} followed by a description of the different magnetic-field scenarios considered here in Section \ref{sec:bfields}.
We focus on magnetic fields close to the source, namely the coherent magnetic field in jet of a BL Lac-type object, as well as the 
turbulent field of a galaxy cluster in which the source might be embedded. 
In Section \ref{sec:method}, we describe the simulated observations and introduce the statistical method.
Most model parameters that influence the strength of the photon-ALP mixing are unknown. 
Therefore, we vary the different parameters
 (ALP mass and coupling, magnetic-field parameters) 
 over large ranges of allowed values and
  investigate the dependence of the sensitivity of a mock IACT on these parameters 
  in  Section \ref{sec:results}
 before concluding in Section \ref{sec:summary}. 
We give a detailed derivation of the turbulent magnetic field in the Appendix.

\section{Photon-ALP mixing}
\label{sec:alps}

The effective Lagrangian for the mixing between photons and ALPs can be written as 
\begin{equation}
\mathcal{L} = \mathcal{L}_{a\gamma} + \mathcal{L}_\mathrm{EH} + \mathcal{L}_a.
\end{equation}
The first term on the right hand side is given in eq. \eqref{eqn:lagr-alps},
the second term is the effective Euler-Heisenberg Lagrangian that accounts for one-loop 
corrections of the photon propagator \cite[e.g.][]{itzykson1984} and the ALP mass and kinetic terms are
\begin{equation}
\mathcal{L}_a = \frac 1 2  \partial_\mu a \partial^\mu a - \frac 1 2 m^2_a a^2.
\end{equation}
ALPs only couple to photons in the presence of a magnetic field component $\mathbf{B}_\perp$ transversal to the propagation direction $\mathbf k$ and only to photon polarisation states 
in the plane spanned by $\mathbf{B}$ and $\mathbf k$
\cite{raffelt1988,deangelis2011}. 
Let $x_3$ be the propagation direction, $\mathbf{B}_\perp = B \hat{\mathbf{e}}_2$, and $A_1$, $A_2$ the polarisation states along $x_1$ and $x_2$, respectively. 
Then the equations of motion for a polarised photon beam of energy $E$ propagating in a cold plasma filled with a homogeneous magnetic field
 read
 \begin{equation}
\left( i\frac{\mathrm{d}}{\mathrm{d}x_3} + E + \mathcal{M}_0 \right)\Psi(x_3) = 0,
\label{eqn:eom}
\end{equation}
with $\Psi(x_3) = (A_1(x_3),A_2(x_3),a(x_3))^T$ and the mixing matrix $\mathcal{M}_0$ (neglecting Faraday rotation),
\begin{equation}
\mathcal{M}_0 = 
\begin{pmatrix}
\Delta_{\perp} & 0 & 0\\
0 & \Delta_{||} & \Delta_{a\gamma} \\
0 & \Delta_{a\gamma} & \Delta_a
\end{pmatrix}.
\end{equation}
The terms $\Delta_{||,\perp}$ arise due to the effects of the propagation of photons in a plasma and the QED vacuum polarisation effect,
$\Delta_{\perp} = \Delta_\mathrm{pl} + 2\Delta_\mathrm{QED}$,  and $\Delta_{||} = \Delta_\mathrm{pl} + 7/2\Delta_\mathrm{QED}$.
The plasma contribution depends on electron density $n_{\mathrm{cm}^{-3}} = n / \mathrm{cm}^{-3}$ through 
the plasma frequency $\omega_\mathrm{pl} \sim 0.037 \sqrt{n_{\mathrm{cm}^{-3}}}\,$neV:
$\Delta_\mathrm{pl} = -\omega_\mathrm{pl} / (2 E)$.
 The QED vacuum polarisation term reads $\Delta_\mathrm{QED} = \alpha E / (45\pi)(B /(B_\mathrm{cr}))^2$, with the fine-structure constant $\alpha$, and the critical magnetic field $B_\mathrm{cr} = m^2_e / |e| \sim 4.4\times10^{13}\,$G.
The kinetic term for the ALP is $\Delta_a = -m_a^2 / (2E)$ and photon-ALP mixing is the result of the off-diagonal elements $\Delta_{a\gamma} = g_{a\gamma} B / 2$. Suitable numerical values for the different $\Delta$ terms are provided in ref. \cite{horns2012icm}.
If photons are lost due to absorption, the diagonal terms $\Delta_{||,\perp}$ get an additional imaginary 
contribution that scales with the mean free path of the photon.
The equations of motion lead to photon-ALP oscillations \cite{raffelt1988,csaki2003,deangelis2007,mirizzi2007,hooper2007:alps} with the wave number
\begin{equation}
\Delta_\mathrm{osc} = \sqrt{(\Delta_{||} - \Delta_a)^2 + 4\Delta_{a\gamma}^2}.
\label{eqn:dosc}
\end{equation}
The conversion probability becomes maximal and independent of energy (the so-called strong mixing regime, SMR) 
for energies $E_\mathrm{crit} \lesssim E \lesssim E_\mathrm{max}$, with \cite[e.g.][]{hooper2007:alps,bassan2009}
\begin{eqnarray}
E_\mathrm{crit} &=& \frac{|m_a^2 - \omega_\mathrm{pl}^2|}{2g_{a\gamma}B} \sim 2.5\,\mathrm{GeV}\, |m^2_\mathrm{neV} - 1.4\times10^{-3}\, n_{\mathrm{cm}^{-3}}| \,g_{11}^{-1} B_{\mu\mathrm{G}}^{-1},
\label{eqn:ecrit}\\
E_\mathrm{max} &=& \frac{90\pi}{7\alpha}\frac{B_\mathrm{cr}^2\,g_{a\gamma}}{B} \sim 2.12\times10^{6}\,\mathrm{GeV}\,g_{11}B_{\mu\mathrm{G}}^{-1}.
\label{eqn:emax} 
\end{eqnarray}
Above $E_\mathrm{max}$, the oscillations are damped due to the QED vacuum polarisation. 
In the above equations, we have introduced the notation $B_X = B / X$, $m_X = m_a / X$, and $g_X = g_{a\gamma} \times 10^X / \mathrm{GeV}^{-1}$.

For an unpolarised photon beam, the problem has to be reformulated in terms of the density matrix $\rho(x_3) = \Psi(x_3)\Psi(x_3)^\dagger$ that obeys the von-Neumann-like commutator equation
\begin{equation}
i\frac{\mathrm{d}\rho}{\mathrm{d}x_3} = [\rho,\mathcal{M}_0],
\end{equation}
which is solved through $\rho(x_3) = \mathcal{T}(x_3,0; E)\rho(0)\mathcal{T}^\dagger(x_3,0; E)$, with the transfer matrix 
$\mathcal T$ that solves eq. \eqref{eqn:eom} with $\Psi(x_3) = \mathcal{T}(x_3,0;E)\Psi(0)$ and initial condition 
$\mathcal{T}(0,0;E) = 1$ \cite[e.g.][]{csaki2003,sanchezconde2009,bassan2010,deangelis2011}. In general, $\mathbf B_\perp$ will not be aligned along $x_2$ but will form an angle $\psi$ with it. 
In this case, the solutions have to be modified with a similarity transformation and, consequently, $\mathcal M$ and $\mathcal T$ will depend on $\psi$ \cite[see e.g.][]{mirizzi2009,bassan2010}. For the mixing in $N_d$ consecutive magnetic domains one finds that the 
photon survival probability of an initial polarisation $\rho(0)$ is given by
\begin{equation}
P_{\gamma\gamma} = \mathrm{Tr}\left( (\rho_{11} + \rho_{22}) \mathcal{T}(x_{3,N_d},x_{3,1};\psi_{N_d},\ldots,\psi_1;E) \rho(0)
\mathcal{T}^\dagger(x_{3,N_d},x_{3,1};\psi_{N_d},\ldots,\psi_1;E)\right),
\label{eqn:surv-prob}
\end{equation}
with $\rho_{11} = \mathrm{diag}(1,0,0)$, $\rho_{22} = \mathrm{diag}(0,1,0)$, and 
\begin{equation}
\mathcal{T}(x_{3,N_d},x_{3,1};\psi_{N_d},\ldots,\psi_1;E) = \prod\limits_{i = 1}^{N_d} \mathcal{T}(x_{3,i+1},x_{3,i};\psi_{i};E).
\end{equation}
For an initially unpolarised photon beam one has $\rho(0) = 1/2\,\mathrm{diag}(1,1,0)$.
The full expression for $\mathcal{T}$ valid at all energies (not only in the SMR) can be found e.g. in ref. \cite{meyer2013thesis}.
For a propagation over a distance $r$ over many domains of coherence length $L_\mathrm{coh}$,
 each with the same constant magnetic field strength $B$
but a random angle $\psi$ that changes from one cell to the next, the average photon-ALP oscillation probability is 
\cite{grossman2002,mirizzi2008}
\begin{equation}
P_{a\gamma} = \frac 1 3\left(1 - \exp\left(-
\frac{3}{2}
\Delta_{a\gamma}^2 r L_\mathrm{coh}\right)\right).
\end{equation}
The more photons convert into ALPs close to the source, the stronger the {\gr} flux enhancement can be as ALPs may reconvert 
into photons in the magnetic field of the Milky Way. 
This leads to a condition for a sufficient mixing over many domains, namely if the absolute value of the argument of the exponent 
in the above equation becomes $\gtrsim 1$. In suitable units this reads
\begin{equation}
g_{11}^2\, B_{\mu\mathrm{G}}^2 \, r_\mathrm{kpc} \, L_\mathrm{kpc} \gtrsim 
2900.
\label{eqn:effic-mix}
\end{equation}
In the case that no photons are absorbed during propagation, the transfer matrix is strictly unitary
and it is easy to show that for an initially unpolarised photon beam the oscillation probability 
is always $P_{a\gamma} \leqslant 1 / 2$ (as shown explicitly in ref. \cite{meyer2013thesis}).

We choose the fiducial ALP parameters to be $m_\mathrm{neV} = 1$ and $g_{11} = 2$.
The choice for the coupling is motivated by the lower limits derived in ref. \cite{meyer2013}
for which photon-ALP mixing could explain indications for a low opacity for VHE $\gamma$ rays.
Neglecting energy and momentum transfer to and from the external magnetic field, such light ALPs are ultra-relativistic with a  $\beta$ factor of
\begin{equation}
\beta = \frac{p_a}{E_a} = \frac{E}{\sqrt{E^2 + m_a^2}} = \left[1 + (m_a / E)^2\right]^{-1/2}
 \approx 1 - \left(\frac{m_a}{2E}\right)^2,
\end{equation} 
with the ALP energy $E_a$, photon energy $E$, and ALP momentum $p_a = E$.
Thus, photon-ALP oscillations will have no effect on the time or spatial structure of the signal. 

\section{Magnetic field scenarios}
\label{sec:bfields} 
Magnetic fields are ubiquitous along the line of sight from the {\gr} source to Earth. 
Here, we consider only those fields for which independent measurements exist, namely the 
fields in the AGN jet, in a galaxy cluster in which the blazar might be embedded, and the magnetic field of the Milky Way (Galactic magnetic field, GMF). 
The contributions to photon-ALP mixing from the $B$ fields of the host galaxy and the intergalactic medium are neglected. 
Typically, BL Lac objects 
and flat spectrum radio quasars (FSRQs)
 are hosted in elliptical galaxies \cite[e.g.][]{matthews1964}. 
While there is strong evidence of micro Gauss fields in these system, little is known about their spatial structure. 
The fields are believed to be turbulent with coherence lengths of the order of 0.1 to 0.2\,kpc  \cite{moss1996} and lacking 
a large scale coherent component in contrast to spiral galaxies \cite[e.g.][]{widrow2002}. 
From eq. \eqref{eqn:effic-mix} one finds that for typical sizes of ellipticals of $r = 10$\,kpc,
an efficient photon-ALP mixing can occur for $g_{11} < 8.8$ only for strong fields, $B \gtrsim 4 \,\mu$G. 
Thus, we ignore this magnetic field. 
For the intergalactic magnetic field (IGMF), only upper limits exist which are of the order of a few $10^{-9}$\,G \cite[e.g.][]{kronberg1994,blasi1999}.
Large scale structure formation simulations and simulations of the deflection of cosmic rays
suggest lower values between $10^{-11}$ to $10^{-12}$\,G \cite{sigl2004,dolag2005}.
Galactic winds, on the other hand, could magnetise the intergalactic medium with field strengths 
$10^{-12} < B < 10^{-8}\,\mathrm{G}$ \cite{bertone2006}
and evidence exists for a redshift evolution of the rotation measure of radio sources
compatible with an IGMF  field strength of 1\,nG and coherence length of 1\,Mpc \cite{neronov2013}.
Only a field strength 
of this value and
 close to the current upper limits and coherence lengths of the order of Mpc will give a sizeable effect on {\gr} spectra \cite[e.g.][]{mirizzi2007,deangelis2007,sanchezconde2009,deangelis2011,dominguez2011alps}. 
Because of the large uncertainties of the intergalactic magnetic-field parameters and the possibility of very low 
$B$-field values (see also the review of ref. \cite{durrer2013}),
we will not consider it here. 

In the following, the adopted models for the $B$-fields in the AGN jet and the galaxy cluster will be described in detail. 
The Galactic magnetic field will be described with the model of ref. \cite{jansson2012}, already used in refs. \cite{horns2012,meyer2013,wouters2014}
to study photon-ALP oscillations. 
The model is derived from a $\chi^2$ fit to WMAP7 synchrotron maps and rotation measures of extragalactic sources. 
In addition to a disk and halo component, it comprises a third out-of-plane component that leads 
to a large photon-ALP conversion probability compared to previous models such as the one provided in ref. \cite[][]{pshirkov2011}.

\subsection{Galaxy clusters}
Evidence exists that a fraction of blazars is harboured in
 galaxy groups or (poor) galaxy clusters \cite[e.g.][]{pesce1995,miller2002,auger2008}.
The existence of turbulent magnetic fields with strength of the order of $\mathcal{O}(\mu\mathrm{G})$ in the intra-cluster medium (ICM) is well established through Faraday rotation measurements and non-thermal (synchrotron) emission at radio frequencies \cite[e.g.][for reviews]{govoni2004,feretti2012}. Evidence exists that the $B$ field follows the electron density in the ICM,
\begin{equation}
B^\mathrm{ICM}(r) = B_0^\mathrm{ICM}(n_\mathrm{el}^\mathrm{ICM}(r) / n_0^\mathrm{ICM})^{\eta_\mathrm{ICM}},
\label{eqn:bicm}
\end{equation}
with $0.5 \lesssim \eta \lesssim 1$, where the electron density can be modelled as 
\begin{equation}
n^\mathrm{ICM}(r) = n_0^\mathrm{ICM}(1 + r/r_\mathrm{core})^{-3\beta_\mathrm{ICM} / 2},
\label{eqn:nicm}
\end{equation}
where typically $\beta_\mathrm{ICM} = 2/3$ and $n_0^\mathrm{ICM} = 10^{-3}\,\mathrm{cm}^{-3}$. 
For many AGN the exact environment is unknown. 
However,  local and intermediate redshift radio galaxies 
are often located  galaxy groups and poor clusters
\cite[e.g.][]{miller2002,auger2008}. 
Therefore, we will look at small clusters with a 
fiducial cluster radius of $r_\mathrm{max} = 300\,$kpc.
 For distances $r > r_\mathrm{max}$ the magnetic field is set to zero. 
 We furthermore assume values for 
 the remaining parameters that are consistent
 with observations of the nearby radio galaxies 3C\,31 \cite{laing2008} and 3C\,449\,\cite{guidetti2010} 
 that reside in poor clusters or galaxy groups, namely $r_\mathrm{core} = 100\,$kpc, $\eta_\mathrm{ICM} = 1$, and $B_0^\mathrm{ICM} = 1\,\mu$G.
In Sec. \ref{sec:results} we will investigate central magnetic 
between $0.1\,\mu\mathrm{G}$ (as found around NGC\,0315 which is situated in a galaxy poor environment \cite{laing2006}) and $10\,\mu\mathrm{G}$ as determined for richer environments such as Hydra\,A \cite[e.g.][]{kuchar2011}.

In terms of photon-ALP oscillations, the turbulent field is often described with a simple cell-like morphology, where the field strength is constant in each cell of length $L_\mathrm{coh}$ but the angle $\psi$ between the magnetic field and the $A_2$ component changes randomly from one cell to the next. 
The coherence length is usually assumed to be of the order of the size of a galaxy in the cluster, i.e. $L_\mathrm{coh} \sim 10\,$kpc.
However, a more physical ansatz for $B$ fields in the ICM is a divergence-free homogeneous and isotropic gaussian turbulent magnetic field with zero mean and variance $\mathcal{B}^2$, that better describes the small and large scale fluctuations seen in observations \cite{murgia2004}.
We will assume here that the power spectrum of the turbulence follows a power law in wave numbers, $M(k) \propto k^q$ between $k_L \leqslant k \leqslant k_H$ and zero otherwise. 
For photon-ALP mixing, only the component transversal to the propagation direction ($x_3$) contributes, where
$B_\perp(x_3) = \sqrt{B^2_1(x_3) + B^2_2(x_3)}$ and $\tan\psi = B_1 / B_2$.
As shown in Appendix \ref{app:bfield}, the transversal components, $i = 1,2$, are given by
\begin{equation}
B_i(x_3) \approx \sum_{n=1}^{N_k}\sqrt{\frac{2\tilde\epsilon_\perp(k_n)\Delta k_n}{\pi}\ln\left(\frac{1}{U_{i,n}}\right)}\cos(k_nx_3+2\pi V_{i,n}),\label{eqn:bfield}
\end{equation} 
where $(U_{i,n}, V_{i,n})$ are uniformly distributed random numbers in the open interval $[0;1)$
and $N_k$ is the total number of spacings in $k$.
The Fourier transform of the correlation function of the transversal $B$-field components is 
\begin{equation}
\tilde\epsilon_\perp(k) = \frac{\pi\mathcal{B}^2}{4} F_q(k ; k_L, k_H),
\label{eqn:eps-trans}
\end{equation}
and $F_q(k; k_L,k_H)$ is given in eq. \eqref{eqn:Fx}. 
We choose an equidistant logarithmic spacing of the $k$ interval where we demand that $N_k \geqslant 10N_D$, where  $N_D$  is the number of decades spanned by the $k$ interval. This ratio is suggested in ref. \cite{kneller2013} in which the authors study neutrino flavour oscillations in turbulent density profiles of supernovae.
Here we set $N_k =\lceil20\log_{10}(k_H / k_\mathrm{min}) \rceil$, with $k_\mathrm{min} = 10^{-3}k_L$.
With this choice, we find a good agreement between the analytical correlation function and simulations (see Appendix \ref{app:bfield}).
The minimum wave number $k_L$ corresponds to the largest scales of energy injection and energy is cascaded down via turbulent fluctuations to  
 the smallest scales set by $k_H$ where  energy dissipation sets in (in turbulent fluids due to viscosity) \cite[e.g.][]{majda1999}.
The model parameters $\mathcal{B}, q, k_L,$ and $k_H$ are a priori unknown but can be constrained through Faraday rotation measurements
at radio frequencies (see e.g. refs. \cite{bonafede2010,kuchar2011} for deduced parameters for the Coma and Hydra A clusters\footnote{
See also ref. \cite{angus2013} for a modelling of the magnetic fields in the Coma cluster including ALPs in order to explain the soft X-ray excess.
}).
For definiteness, we will use the magnetic-field spectrum derived for Coma \cite{bonafede2010} 
and set $k_L = 0.18\,\mathrm{kpc}^{-1}$, $k_H = 3.14\,\mathrm{kpc}^{-1}$, and $q = -11/3$,
which is in accordance with the values deduced for the environment of 3C\,31 \cite{laing2008}. 
The value of $q$ corresponds to a Kolmogorov-type power spectrum, so that the magnetic energy, $u(k) = 4\pi k^2 M(k)$, follows a power law with index $-5/3$.
The choice for $k_H$ ensures that the maximum wave number is larger than the oscillation wave number, $k_H > \Delta_\mathrm{osc}$, given in eq. \eqref{eqn:dosc}, for the fiducial magnetic field  $\mathcal{B} = 1\,\mu$G and $g_{11} = 2$, since 
at the critical energy one finds $\Delta_\mathrm{osc}\sim0.04\, g_{11} B_{\mu\mathrm{G}}$.
For the numerical calculations, we also have to set the integration step length, $L$. 
In order to take the smallest scale for fluctuations into account, we choose $L = 1 / k_H$.
Finally, inserting $B_\perp$ for $B_0^\mathrm{ICM}$ in eq. \eqref{eqn:bicm} accounts 
for the radial decrease of the magnetic-field strength. 

\subsection{BL Lac jets}
Magnetic fields in blazar jets have been deduced from measurement at different scales, 
ranging from ultra-compact regions at distances of $\sim 0.1$pc from the central black hole up 
to 100\,kpc scale structures such as lobes, plumes, and hot spots \cite{pudritz2012}. 
Radio observations of the frequency dependent positional shifts of the optically thick radio cores 
allow the estimation of the core magnetic-field strength, usually of the order of 0.1\,G \cite{osullivan2009} at a distance of 1\,pc to the central black hole\footnote{
The studied source sample of ref. \cite{osullivan2009} consists of six objects, four of which are classified as BL Lacs.
}. 
These values are of the same order as the ones inferred from the modelling of spectral energy distributions (SEDs) from 
simple synchrotron-self-compton (SSC) emission models \cite[e.g.][]{celotti2008,tavecchio2010:BLLac} that probe the 
magnetic field at the VHE emission zone. In these models, however, the emission sites are located at distance $< 1\,$pc (see below).

We focus here on BL Lac-type objects. In this case, we can neglect internal {\gr} absorption and photon-ALP oscillations within the broad line
region (BLR) \cite{tavecchio2014}, since, by definition, these sources show either no or only weak emission lines 
in their optical spectra. 
We model the photon-ALP conversion in the parsec-scale jet similar to the prescriptions outlined in ref. \cite{tavecchio2012,mena2013,tavecchio2014}.
Evidence exists that the magnetic field in the BL Lac jet can be modelled with a poloidal (along the jet-axis) and a toroidal (perpendicular to the jet axis) coherent component, where the field strength for the former decreases as $B \propto r^{-2}$ and $B \propto r^{-1}$ for the latter
\cite{begelman1984,rees1987,pudritz2012}. 
Consequently, for large enough distances to the central black hole, the toroidal component dominates and we neglect the poloidal component. 
Assuming equipartition between the magnetic and particle energies, the electron densities typically range from $\sim100$ to $\sim1000\,\mathrm{cm}^{-3}$ at 1\,pc from the AGN core and the electron density profile follows $n_\mathrm{el} \propto r^{-2}$  \cite[][]{osullivan2009}.
 Accordingly, we adopt the following prescriptions for the magnetic field and the electron density of the parsec scale jet 
\begin{eqnarray}
B^\mathrm{jet}(r) &=& B^\mathrm{jet}_0 \left(\frac{r}{r_\mathrm{VHE}}\right)^{\eta_\mathrm{jet}},\\
n^\mathrm{jet}_\mathrm{el}(r) &=& n^\mathrm{jet}_0 \left(\frac{r}{r_\mathrm{VHE}}\right)^{\beta_\mathrm{jet}}.
\end{eqnarray}
with $\eta_\mathrm{jet} = -1$, $\beta_\mathrm{jet} = -2$, and $r_\mathrm{VHE}$ the distance of the VHE emission site to the central black hole. 
Assuming $r_\mathrm{VHE} \sim R_\mathrm{VHE} / \theta_\mathrm{j}$, where $R_\mathrm{VHE}$ is the radius of the VHE emitting plasma blob
and $\theta_\mathrm{j}$ is the angle between the jet axis and the line of sight, typical values for $r_\mathrm{VHE}$ 
range from $\sim 0.01$\,pc to $\sim0.1$\,pc, deduced from SSC models \cite{celotti2008}. 
Our fiducial choice is $r_\mathrm{VHE} = 0.01\,$pc. 
The magnetic field is supposed to be coherent, so $\psi = 0$ is chosen over the entire jet region. 
As noted in ref. \cite{tavecchio2014}, the above equations hold in the co-moving frame of the jet. The photon energy $E^\prime$ in this frame 
is related to the energy $E$  in the laboratory frame through the Doppler factor, $E^\prime = E / \delta_\mathrm{D}$,
where $\delta_\mathrm{D} = [\Gamma_\mathrm{L}( 1 - \beta_\mathrm{j}\cos\theta_\mathrm{j})]^{-1}$ with the relativistic Lorentz and beta
factors $\Gamma_\mathrm{L},\beta_\mathrm{j}$ of the bulk plasma movement, respectively.
The photon-ALP conversion is numerically calculated following eq. \eqref{eqn:surv-prob}.
For the numerical calculations, the step length along $x_3$ is chosen such that the magnetic field 
decreases (with $r^{-1}$) by one percent from one cell to the next. We consider the coherent magnetic field up to $r_\mathrm{max} = 1\,$kpc. For our fiducial scenario, the electron density is set to  $n_0^\mathrm{jet} = 10^4\,\mathrm{cm}^{-3}$ and the magnetic-field strength is set to $B_0^\mathrm{jet} = 0.1$\,G. We choose $\theta_\mathrm{j} = 1^\circ$ and $\Gamma_\mathrm{L} = 8$ leading to $\delta_\mathrm{D} \sim 15.64$. 
We underline that in this scenario we are only taking the jet $B$-field and the GMF into account and
 we apply a fully energy-dependent modelling of the photon-ALP mixing
so that no restriction to the SMR is necessary.

As mentioned above, magnetic fields are also present at kpc scales of the jet, especially in jet features 
of FSRQs such as hot spots and lobes,
with magnetic fields up to $\mathcal{O}(100\,\mu\mathrm{G})$ in the former and $\mathcal{O}(1\,\mu\mathrm{G})$ 
in the latter \cite{pudritz2012}.
It is noteworthy that the field strength is comparable to the one found in galaxy clusters and thus
these sites are of potentially high interested for photon-ALP conversions, as pointed out in ref. \cite{tavecchio2014}.
Interestingly, Faraday rotation observations also suggest the existence of turbulent magnetic fields
 in the radio lobes of the misaligned BL Lac \cite{chiaberge2001} Centaurus A
 with $B \sim 0.8\,\mu$G, a coherence length of $L_\mathrm{coh} \sim 20\,$kpc and a total size of the region of $\sim180$\,kpc \cite{feain2009}
  Thus, if similar magnetic-field morphologies for the large scale jet and a galaxy cluster are assumed, the results for 
a small-scale galaxy cluster should also be comparable to that of a large-scale AGN jet structure.

\begin{figure}
\centering
\includegraphics[width = .7 \linewidth]{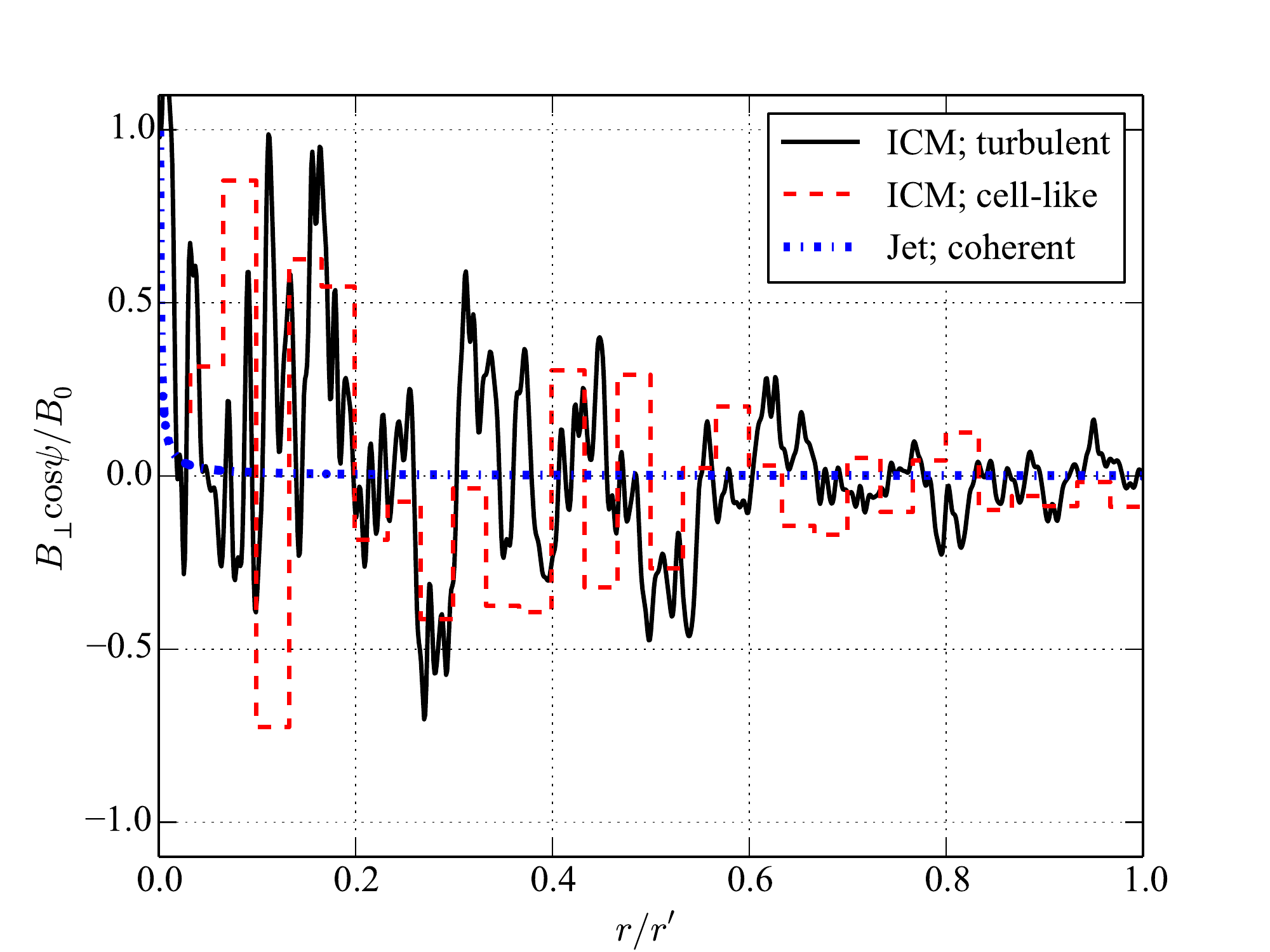}
\caption{Tested magnetic-field scenarios, with the fiducial parameters of table \ref{tab:fields}. For the ICM fields, 
one particular realisation is shown. In the ICM case $r^\prime = r_\mathrm{max}$ 
whereas $r^\prime = 40\,$pc in the jet scenario for better visibility. The fields are normalised to their maximum values and 
multiplied with $\cos\psi$ which gives the field strength along the $A_2$ component.}
\label{fig:fields}
\end{figure}

In summary, three magnetic-field scenarios will be compared: The coherent parsec-scale field in the BL Lac jet, and the magnetic field in a galaxy cluster modelled with two different morphologies. The three magnetic fields normalised to their maximum values are shown in figure \ref{fig:fields} and the fiducial parameter values are summarised in table \ref{tab:fields}.
The radial decrease in all scenarios is evident. 
  
\begin{table}[t!b]
\centering
\begin{tabular}{|l|ccc|}
\hline
{} & \multicolumn{3}{c|}{$B$-field scenarios} \\
Parameter & AGN jet & ICM, cell-like & ICM, turbulent \\
\hline
$B_0\,(\mu\mathrm{G})$ & $10^5$ & $1$ & $1$ \\
$r_\mathrm{VHE}\,(\mathrm{pc})$ & 0.01 & -- & -- \\
$r_\mathrm{max}$ (kpc) & 1 & 300 & 300 \\
$\delta_\mathrm{D}$ & 15.64 & -- & -- \\
$L\,$(kpc) & adaptive & 10 & $1 / k_H$\\
$\eta$ & $-1$ & 1 & 1\\
$\beta$ & $-2$ & $2/3$ &  $2/3$ \\ 
$r_\mathrm{core}$ (kpc) & -- & 100 & 100 \\
$n_0\,(\mathrm{cm}^{-3})$ & $10^4$ & $10^{-3}$ & $10^{-3}$\\ 
$q$ & -- & -- & $-11/3$ \\
$k_L\,(\mathrm{kpc}^{-1})$ & -- & -- & $0.18$ \\
$k_H\,(\mathrm{kpc}^{-1})$ & -- & -- & $3.14$ \\
$N_k$ & -- & -- & $\lceil20(\log_{10}(k_H / k_L) +3)\rceil $ \\
\hline
\end{tabular}
\caption{Fiducial model parameters for the different magnetic fields. The step length for the cell-like model 
is equal to the coherence length in that model. 
See text for further details.}
\label{tab:fields}
\end{table}

\begin{figure}
\centering
\includegraphics[width = .7 \linewidth]{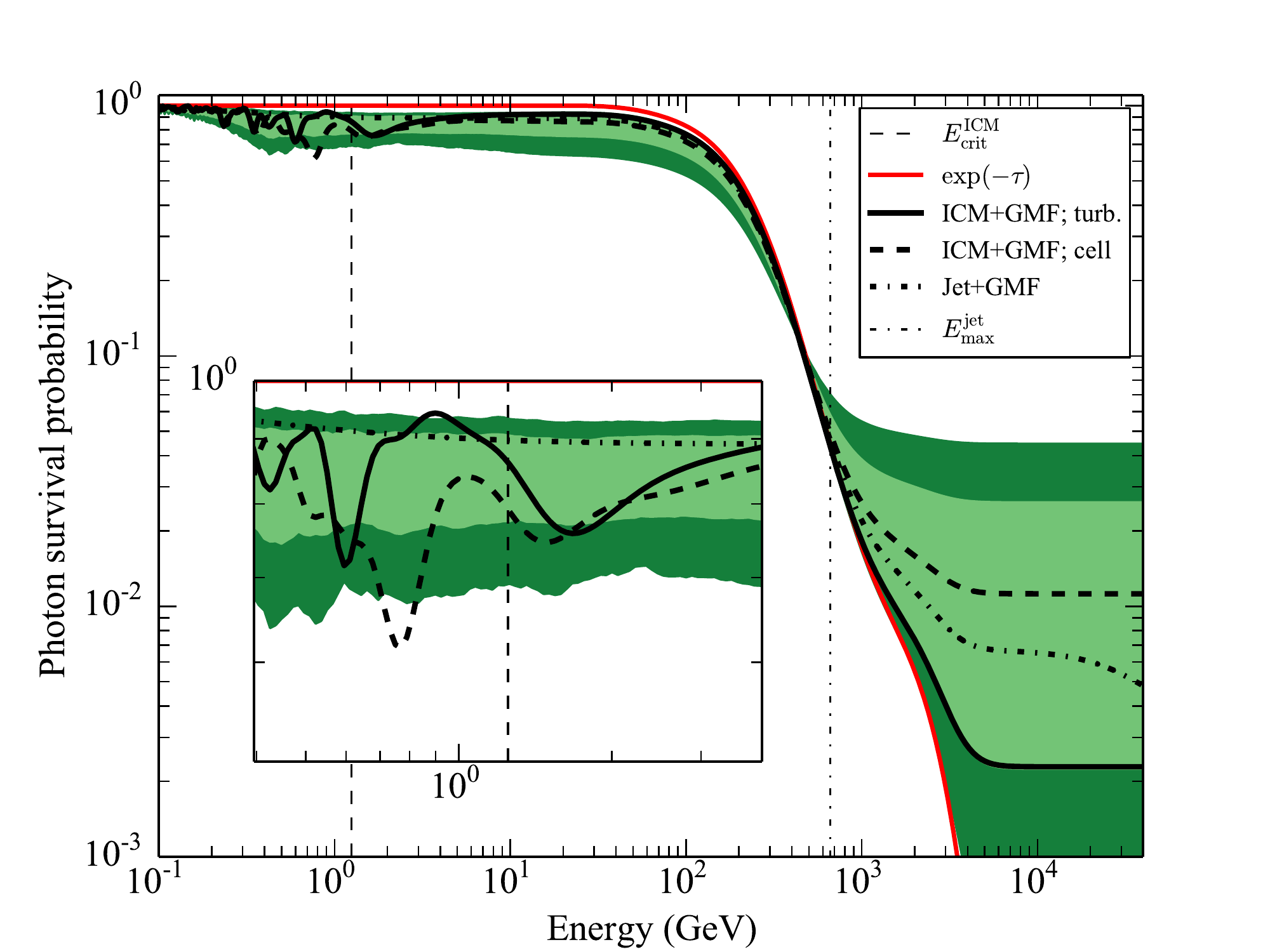}
\caption{Photon survival probability for the different magnetic field scenarios. The sky position of the blazar PG\,1553+113 is assumed 
with $z = 0.4$. The envelopes refer to the turbulent $B$-field scenario, only.
They show the 68\,\% (light green) and 95\,\% (dark green) contours around the median.
 For the ICM scenarios, the black lines show the result of one random $B$-field realisation, the red line indicates the attenuation in the absence of ALPs.
  The inset shows a zoom-in on the energy regime around $E_\mathrm{crit}$ of the ICM scenarios. Above $E_\mathrm{max}^\mathrm{jet}$ (in the lab frame) for the jet scenario, the QED effect sets in leading to oscillations in $P_{\gamma\gamma}$. The fiducial parameter values of table \ref{tab:fields} are used, together with $g_{11} = 2$ and $m_\mathrm{neV} = 1$. }
\label{fig:pgg}
\end{figure}

The photon survival probability for the three models is shown in figure \ref{fig:pgg} for a source at $z = 0.4$ and sky coordinates 
coincident with the position of the blazar PG\,1553+113 (see Section \ref{sec:method}). 
Throughout this article, the EBL model introduced in ref. \cite{kneiske2010}
 which predicts a minimum absorption for $\sim$\,TeV $\gamma$ rays is applied. 
For the random ICM fields, 1000 realisations are simulated. The green envelopes show the regions 
in which 68\,\% (95\,\%) of all realisations around the median of the gaussian turbulent $B$-field are contained.  
In all scenarios most $B$-field realisations lead to a boost in {\gr} flux beyond 
an optical depth of $\tau \gtrsim 2$. 
Above the critical energy (shown by the dashed line for the ICM case),
the photon-ALP mixing enters the SMR leading to a drop in the survival probability.
Around $E_\mathrm{crit}$, an oscillatory behaviour of $P_{\gamma\gamma}$ 
is observed that has been used to constrain $m_a$ and $g_{a\gamma}$ \cite{ostman2005,wouters2012,hess2013:alps}.

In the next section, we introduce a method to quantify the sensitivity to detect photon-ALP signatures and the 
influence of the different parameters on the sensitivity is discussed in Section \ref{sec:results}.

\section{Method}
\label{sec:method}

With the magnetic field models at hand, we now investigate the impact of photon-ALP oscillations on an AGN spectrum. 
This will be done by generating a mock data set of a hypothetical observation of a blazar with a CTA-like array (Section \ref{sec:obs-irf}).
In Section \ref{sec:stats}, we introduce the statistical test with which we quantify the sensitivity of the array to detect the ALP signal.

\subsection{Intrinsic and observed blazar spectrum}
\label{sec:obs-irf}
Blazars are known to be time variable and episodes of increased {\gr} activity offer the opportunity to obtain 
high signal-to-noise spectra at large optical depths. 
Therefore, we will assume an observation of the BL Lac object PG\,1553+113 
located at a sky position\footnote{The position of the source in the sky determines the reconversion probability in the Galactic magnetic field \cite{simet2008,horns2012icm,wouters2014}.} of $\alpha_\mathrm{J2000} = 15^\mathrm{h}55^\mathrm{m}43.0^\mathrm{s}$ and $\delta_\mathrm{J2000} =+11^\mathrm{d}11^\mathrm{m}24.3^\mathrm{s}$.
A lower limit on the redshift of $z \geqslant 0.4$ has been inferred from Ly-$\alpha$ 
absorption lines in the optical spectrum \cite{danforth2010}.
The source has been observed in the VHE regime with H.E.S.S. \cite{pg1553hess2005,pg1553hess2008}, MAGIC \cite{pg1553magic2007},
and VERITAS \cite{pg1553veritas2010}, and it underwent 
a flaring episode in 2012 where its integrated flux above 100\,GeV reached the level of the Crab nebula \cite{atel4069}.
To illustrate our method, we will make the assumption that this source is located in a galaxy cluster
when considering the ICM scenarios.

The 2005 H.E.S.S. observation of PG\,1553+113 \cite{pg1553hess2008} serves as a template for the intrinsic blazar spectrum. 
The observation covers the energy range 0.23\,TeV to 1.23\,TeV, corresponding to $0.8 \lesssim \tau \lesssim 4.4$  for $z = 0.4$. 
Consequently, the measurement could already suffer from any non-standard photon propagation
and we decide to estimate the intrinsic spectrum from the first three energy bins or up to an optical depth of $\tau \sim 1.86$. 
These bins are corrected for absorption with and without ALPs and fitted with a power law, $\phi(E) = N (E/E_0)^{-\Gamma}$. 
Since the fit comprises only three data points, we always find a $\chi^2$ value over degrees of freedom close to one.
The best-fit power-law normalisation is upscaled by a factor of $C = 3.58$ to emulate the flaring episode of the source,
which is assumed over the entire observation time\footnote{
The factor $C$ is determined by fitting the observed 2005 H.E.S.S. spectrum with a power law in the entire energy range and 
subsequently integrating the spectrum above 100\,GeV in order to determine the flux in units of the Crab nebula (where the
Crab spectrum of ref. \cite{aharonian2006crab} is assumed). In order to reach 100\,\% of the Crab nebula's flux as reported by the MAGIC collaboration in 2012 \cite{atel4069}, the normalisation has to be upscaled by $C$.  
}.
With ALPs, the intrinsic spectrum has to be re-determined for every chosen set of parameters and, in case of the ICM scenarios,
for every random realisation of the magnetic field. 
In future observations with CTA, the intrinsic spectrum can be determined more accurately, due to the 
lower energy threshold. 
It is assumed that the determined power law is a valid description of the intrinsic spectrum up to an optical depth of $\tau = 12$ or an energy of $E \sim 7.4$\,TeV. 
For higher energies, the source flux is set to zero. 

\begin{figure}
\centering
\includegraphics[width = .7 \linewidth]{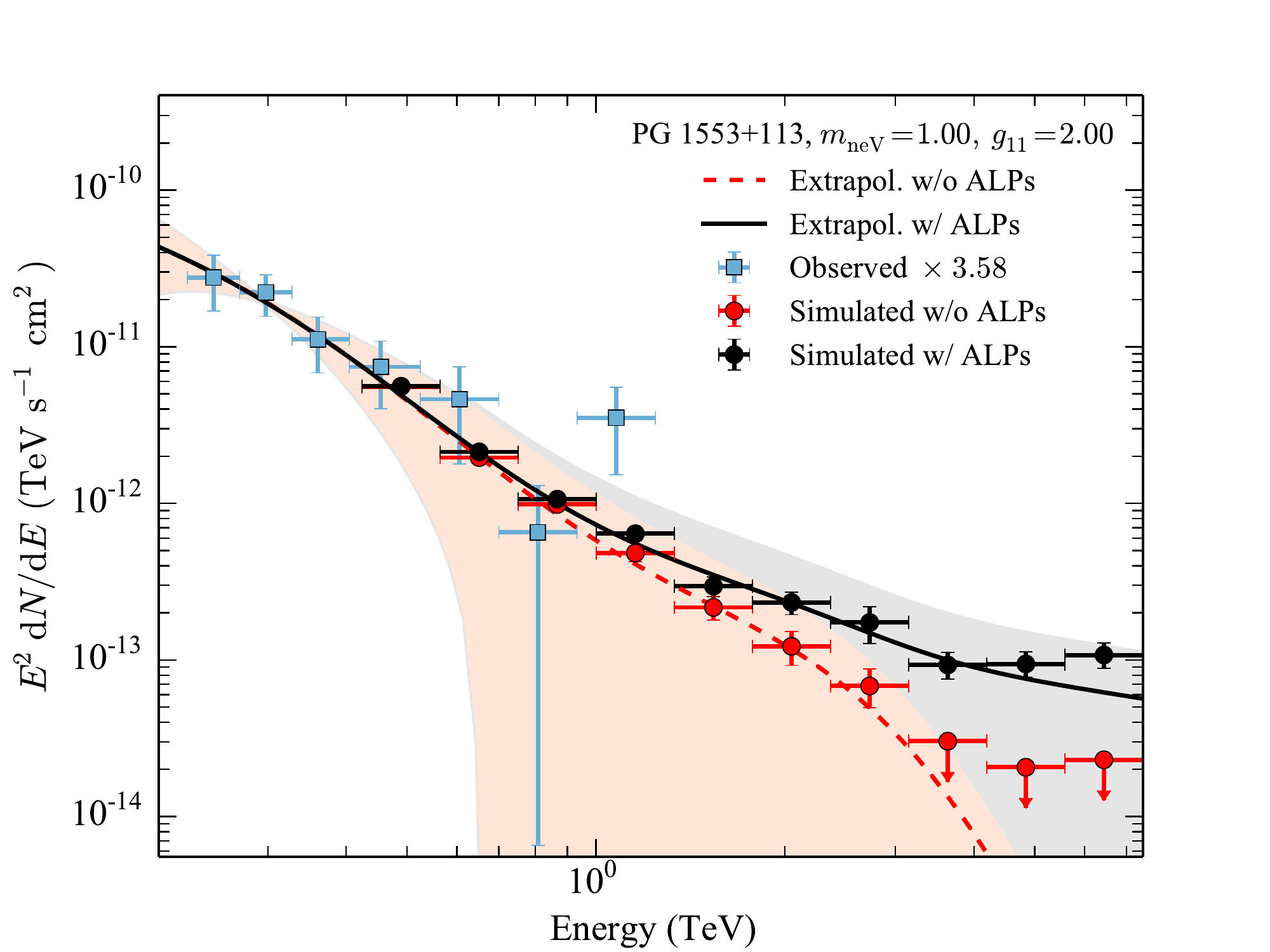}
\caption{Example of a simulated spectrum above $\tau = 1$ with and without an photon-ALP oscillations. The ALP contribution is shown for the fiducial turbulent ICM scenario 
for one random $B$-field realisation and fixed mass and coupling. The extrapolated spectra (dashed and solid lines) are the results of multiplying
the intrinsic best-fit spectrum with the photon survival probability, see eq. \eqref{eqn:spec_obs}. The shaded regions correspond 
to the fit uncertainties of the intrinsic spectrum.
In case the significance in one bin is below $2\,\sigma$, $2\,\sigma$ upper limits on the flux are shown.
 The H.E.S.S. data points are upscaled by 3.58 to emulate the {\gr} flare of PG\,1553+113.}
\label{fig:spec}
\end{figure}

The {\gr} spectrum at Earth is given by 
\begin{equation}
\phi_0(E) = P_{\gamma\gamma} \phi(E),
\label{eqn:spec_obs}
\end{equation}
where $P_{\gamma\gamma} $ reduces to $\exp(-\tau)$ in the no-ALP case, i.e. $g_{11} = 0$.
The expected number of counts is obtained by folding the observed spectrum with the instrumental response function (IRF);
A function of the true energy $E$ and reconstructed energy $E^\prime$. It is assumed that the IRF factorises into 
the point spread function (PSF), the effective area $A_\mathrm{eff}$, and the energy dispersion, $D_E$. 
For the IRF of a CTA-like array we use the published results of CTA Monte-Carlo studies, 
namely the shower axis maximisation (SAM) analysis for the array I configuration, 
provided in ref. \cite{bernloehr2013}, see especially figure 15 for the effective area, figure 17 for the PSF, and figure 18 for the energy resolution. 
In ref. \cite{bernloehr2013} the different components of the IRF are obtained with Monte-Carlo simulations under the assumption of a zenith angle of $20^\circ$ and a ratio between the exposures of background and signal regions of $\alpha = 0.2$, which we also adopt here.
We approximate the energy dispersion with a gaussian function of variance $\sigma_E(E) = 0.1 E$ and neglect the PSF.
The array I configuration is a compromise between a good sensitivity at low and high energies \cite{bernloehr2013}. 
It is therefore suitable for ALP searches, since it guarantees a good determination of the intrinsic spectrum, as well as sufficient 
sensitivity at high energies and correspondingly high optical depths even for close-by AGN. 
 In each energy bin $i$ of width $\Delta E_i^\prime$ the expected number of counts is,
 \begin{equation}
 \mu_i = T_\mathrm{obs} 
 \int_{\Delta E_i^\prime}d E^\prime A_\mathrm{eff}(E^\prime) \int  d E\, D_E(E^\prime,E) \phi_0(E), 
 \end{equation}
for an observational time $T_\mathrm{obs}$ which we set to 20 hours\footnote{
The 2005 data set of the H.E.S.S. observations comprises 7.6\,hours of data \cite{pg1553hess2008}.
}. 
The expected number of residual background events in each bin, $b_i$, is determined by multiplying $T_\mathrm{obs}$ with the background rate,
provided in figure 16 of ref. \cite{bernloehr2013}. 
The number of  events from the source region (ON) and from a background (OFF) region is drawn from the corresponding 
Poisson distributions (suppressing the bin index $i$), 
\begin{eqnarray}
f(N_\mathrm{ON} | \mu + b) &=& \frac{(\mu + b)^{N_\mathrm{ON}}}{N_\mathrm{ON}!}\exp(-(\mu +b)), \label{eqn:fON}\\
f(N_\mathrm{OFF} | b /  \alpha) &=& \frac{(b/\alpha)^{N_\mathrm{OFF}}}{N_\mathrm{OFF}!}\exp(-b/\alpha).\label{eqn:fOFF}
\end{eqnarray}
The number of excess events from the {\gr} source is then $N_\mathrm{excess} = N_\mathrm{ON} - \alpha N_\mathrm{OFF}$ and the signal significance is calculated with eq. (17) of ref. \cite{lima1983}.
The resulting simulated spectra above $\tau = 1$ with and without an ALP contribution are shown in figure \ref{fig:spec} together with 
the H.E.S.S. observation. The secondary flux of regenerated $\gamma$ rays starts to contribute significantly above $\tau \gtrsim 3$ 
and dominates the spectrum above $E \sim 2$\,TeV ($\tau \sim 5.2$). 

\subsection{Statistical test to search for an increased {\gr} flux}
\label{sec:stats}
Our technique to search for an ALP signal at high optical depths is based 
on a likelihood ratio test between the following two hypotheses:  An ALP exists and gives rise to 
a boost of the {\gr} flux and expected number of signal counts $\boldsymbol\mu(g_{a\gamma},m_a, \boldsymbol{\theta}) =
(\mu_1,\ldots,\mu_n)$, with $\mu_i = \mu_i(g_{a\gamma},m_a, \boldsymbol{\theta}),\,i = 1,\ldots,n$, the expected count 
number in each energy bin.
Alternatively, the {\gr} flux is determined by standard EBL absorption only, with expected counts $\tilde{\boldsymbol\mu} \equiv \boldsymbol\mu(g_{11} = 0,m_a,\boldsymbol\theta) = \boldsymbol\mu(g_{a\gamma}, m_\mathrm{neV} \gg 1,\boldsymbol{\theta})$.
The last equality holds since for large ALP masses, the critical energy is shifted to energies inaccessible with IACTs in the considered 
$B$-field scenarios. 
The vector $\boldsymbol{\theta}$ encodes all further nuisance parameters that describe the signal strength, e.g., 
the spectral parameters, which are fixed to the power-law the extrapolation of the intrinsic spectrum.  
It is assumed that the intrinsic spectrum does not harden with energy, as discussed in Section \ref{sec:intro}.
Further nuisance parameters are the magnetic-field parameters and the chosen EBL model,
all fixed here by our model selections. 

Given the observation of $N_\mathrm{ON}$ events from the source region 
and $N_\mathrm{OFF}$ from a background region in $n$ energy bins, 
the likelihood for an expected number of source and background counts
$\boldsymbol{\mu}$, $\mathbf{b}$ is defined as the product of the probability 
mass functions of eq. \eqref{eqn:fON} and \eqref{eqn:fOFF} in all energy bins,
\begin{equation}
\mathcal L (\boldsymbol\mu,\mathbf b;\alpha | N_{\mathrm{ON}}, N_{\mathrm{OFF}})
= \prod\limits_{i\atop\tau(E_i,z) > 2}^n f(N_{i,\mathrm{ON}} | \mu_i + b_i) f(N_{i,\mathrm{OFF}} | b_i /  \alpha).
\label{eqn:likelihood}
\end{equation}
Since we are interested in the ALP effect at high optical depth,
we restrict the likelihood to only include the energy bins 
for which the  central energy $E_i$  fulfils the condition $\tau(E_i,z) > 2$. 
We are using 10 energy bins evenly spaced in logarithmic energy between $\tau = 2$ and $\tau = 12$. 
Only bins 
that are detected with a significance larger than $2\,\sigma$ (using eq. (17) in \cite{lima1983})
 are used. 
Bins at the high-energy end of the spectrum falling below this threshold will be combined and used if the significance 
exceeds the threshold after rebinning.   
We compare the no-ALP hypothesis (with expected counts $\tilde{\boldsymbol{\mu}}$)
with the ALP hypothesis (and expected counts $\boldsymbol{\mu}$) with the 
likelihood ratio test \cite[e.g.][]{rolke2005}
\begin{equation}
\lambda (\tilde{\boldsymbol\mu};\alpha | N_{\mathrm{ON}}, N_{\mathrm{OFF}} )= 
\frac{
\mathcal L (\tilde{\boldsymbol\mu},\widehat{\widehat{\mathbf b}}(\tilde{\boldsymbol\mu}) ;\alpha | N_{\mathrm{ON}}, N_{\mathrm{OFF}})
}
{
\mathcal L (\widehat{\boldsymbol\mu},\widehat{\mathbf b};\alpha | N_{\mathrm{ON}}, N_{\mathrm{OFF}})
}.\label{eqn:loglRatio}
\end{equation}
In the numerator, the likelihood is maximised by $\widehat{\widehat{\mathbf b}}$ for fixed $\tilde{\boldsymbol\mu}$  while 
in the denominator it is maximised with respect to both, $\boldsymbol\mu$ and $\mathbf b$,
 with $\widehat{\boldsymbol\mu}$ and $\widehat{\mathbf b}$ being the maximum-likelihood estimators.
The test statistic,
\begin{equation}
\mathit{TS} = -2\ln\lambda,
\end{equation}
converges to a $\chi_\nu^2$ distribution with $\nu$ degrees of freedom (d.o.f.) that can be used 
to calculate the significance with which the no-ALP hypothesis can be excluded.
Likewise, limits and confidence intervals in the $(m_a,g_{a\gamma})$ plane can be derived by exchanging $\tilde{\boldsymbol\mu}$ with 
${\boldsymbol\mu}$ in eq. \eqref{eqn:loglRatio} and re-evaluating the test statistic and the significance.  

The number of d.o.f. for the $\chi^2_\nu$ distribution (the null distribution) 
to determine the $p$-value with which one can exclude the no-ALP hypothesis 
is a priori unknown.
According to Wilks' theorem \cite{wilks1938}, it should be equal 
to the difference of d.o.f. of the two hypotheses. 
However, the observable (the number of $N_\mathrm{ON}$ counts and with it 
the $\mathit{TS}$ values) does not depend linearly on the parameters as required by Wilks' theorem.
For example, higher values of $g_{11}$ do not always result in a stronger boost due to the dependence on the specific magnetic-field realisation.
Secondly, for certain parameter choices, the ALP mixing vanishes and the remaining model parameters are unconstrained
(e.g. if the coupling is zero, the value of the magnetic field is irrelevant). Thus, it is non-trivial to determine 
the difference in the degrees of freedom between the two hypotheses. 
Therefore,  the null distribution of the $\mathit{TS}$ values is determined via Monte-Carlo simulations.

\begin{figure}
\centering
\includegraphics[width = .49 \linewidth]{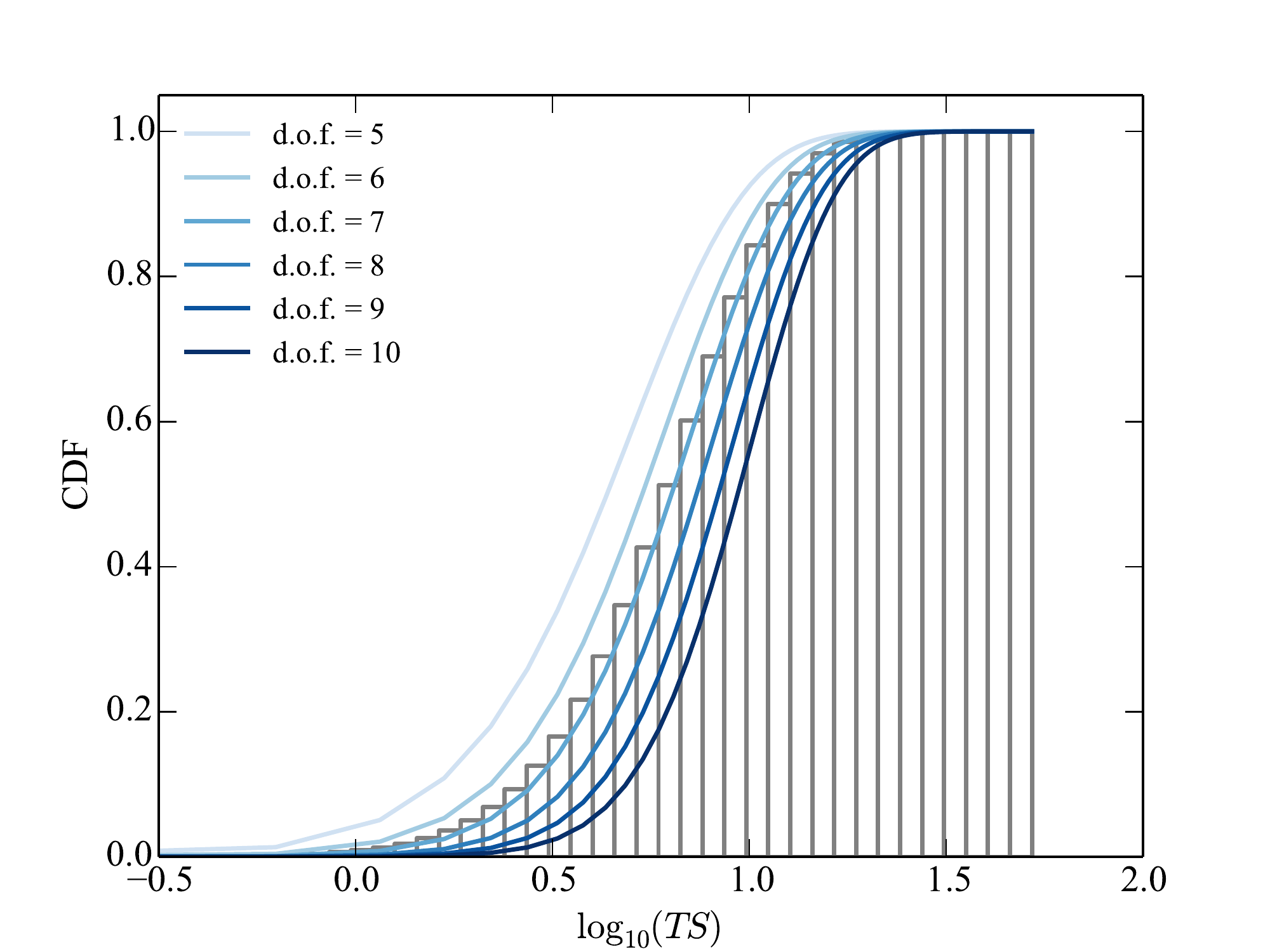}
\includegraphics[width = .49 \linewidth]{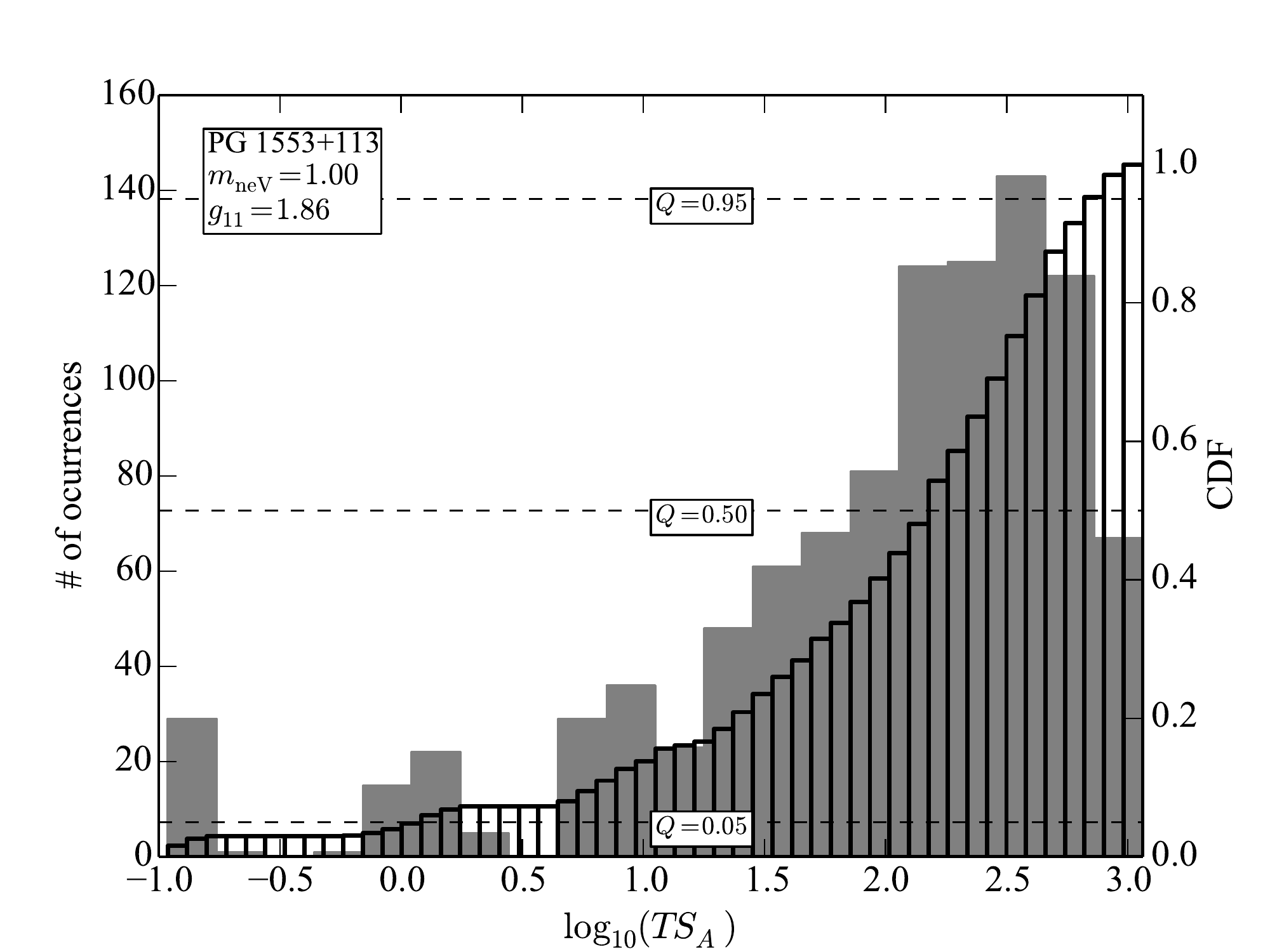}
\caption{$\mathit{TS}$ distributions. \textit{Left:} CDF of the null distribution of the $\mathit{TS}$ values compared 
the $\chi^2$ distributions with different degrees of freedom. \textit{Right:} $\mathit{TS}_A$ distribution of different realisations of  
the turbulent cluster $B$-field for one ALP mass and coupling. 
The grey histogram shows the distribution, the CDF (left $y$-axis) is given by the transparent histogram in the foreground.
The different quantiles $Q$ of the CDF are shown as dashed lines.}
\label{fig:TS}
\end{figure}

We simulate $10^6$ observations for an ALP coupling of $g_{11} = 0$
and calculate the $\mathit{TS}$ values. 
The likelihood in the  denominator of eq. \eqref{eqn:loglRatio} is maximised with 
respect to both $\mu_i$ and $b_i$ with  
$\widehat{\mu}_i = N_{i,\mathrm{ON}} - \alpha N_{i,\mathrm{OFF}} $ and $\widehat{b}_i = N_{i,\mathrm{OFF}}/\alpha$.
The cumulative distribution of the simulations is shown in the left panel of figure \ref{fig:TS} together 
with the cumulative distribution functions (CDF) of $\chi^2$ distributions for different d.o.f.
The distribution is very well described with a $\chi^2_\nu$ distribution with $\nu=6$. 

An efficient way to determine the sensitivity to reject the no-ALP hypothesis without 
the necessity to simulate a large number of observations is the use of an Asimov data set, for 
which the maximum likelihood estimators of the true values are the expected values \cite{cowan2011}.
For such a data set the number of ON and OFF events are equal to the expected values, so for each bin
$N_{i, \mathrm{ON}} = \mu_i + b_i$ and $N_{i, \mathrm{OFF}} = b_i / \alpha$ \cite[][]{cowan2011}. 
The non-integer values are of no concern since we are only interested in the value of the likelihood ratio test, which reads with the Asimov data set
\begin{equation}
\mathit{TS}_A = -2\ln \lambda_A (\tilde{\boldsymbol\mu};\alpha)= -2 \ln\left(
\frac{
\mathcal L (\tilde{\boldsymbol\mu},\widehat{\widehat{\mathbf b}}(\tilde{\boldsymbol\mu}) ;\alpha | \boldsymbol\mu + \mathbf b, \mathbf b / \alpha)
}
{
\mathcal L (\boldsymbol\mu,\mathbf b;\alpha | \boldsymbol\mu + \mathbf b, \mathbf b / \alpha)
}\right).
\end{equation}
The Asimov data set gives an estimation for the median value of the $\mathit{TS}$ distribution which otherwise needs
to be determined from a large number of Monte-Carlo simulations \cite{cowan2011}. 

For the scenarios including mixing in galaxy clusters, 
we simulate 1000 random realisations of the $B$ field 
for each chosen parameter set and compute $\mathit{TS}_A$ for each set. 
An example for the $\mathit{TS}_A$ distribution is shown in right panel of figure \ref{fig:TS}. 
The distribution is highly skewed, and changes if different parameters are chosen.
In the following, 
for each set of parameters $(m_a,g_{a\gamma},\boldsymbol\theta)$, we will consider
specific realisations of the turbulent magnetic field that result in 
 $\mathit{TS}_A$ values that correspond to different quantiles $Q$ of the CDF, 
  $\mathrm{CDF}(\mathit{TS}_A) = Q$. 
For instance, the  realisation giving the $\mathit{TS}_A$ value with $\mathrm{CDF}(\mathit{TS}_A) = 0.5$ 
represents the $B$-field configuration resulting in the median of the $\mathit{TS}_A$ distribution.
From the right panel of figure \ref{fig:TS} 
it is evident that the bulk of $B$-field realisations result in a deformation of the AGN spectrum and high $\mathit{TS}_A$ values. 
However, for some orientations, the spectral changes are marginal. 
For a cell-like structured magnetic field, in the SMR, one can show that for an even number of domains the ALP effect can completely vanish
if one half of the domains forms an angle $\psi + \pi$ and the other half an angle $\psi$ between the $x_2$ component and $\mathbf{B}_\perp$ \cite{meyer2014dom}. 
In the following, the results will be presented for $Q = 0.5$ and $Q = 0.05, 0.95$.  
For $Q = 0.05$, 95\,\% of all realisations give a higher test statistic and the corresponding $B$-field 
can be regarded as pessimistic in terms of photon-ALP mixing. 
In the other case, $Q = 0.95$, only 5\,\% of the simulated turbulent $B$-fields result 
in a higher detection of the spectral deformations. Consequently, this configuration 
is the optimistic case for photon-ALP mixing.

\section{Application to mock data}
\label{sec:results}
We now illustrate the likelihood ratio test and the impact of the modelling of the coherent and turbulent magnetic fields by applying the test to a mock data set of an observation of PG\,1553+113.
The dependence of the test statistic values of the Asimov data, $\mathit{TS}_A$,
to the different model parameters is investigated by stepping through a value range of the parameter in question 
and setting all other parameters to their fiducial values. 
The fiducial values for the $B$ fields are given in table \ref{tab:fields} and the fiducial ALP parameters are $g_{11} = 2$ and $m_\mathrm{neV} = 1$ 
(see Section \ref{sec:alps}).
The coherent Galactic magnetic field is fixed to the model of ref. \cite{jansson2012}.

\subsection{Cluster magnetic-field parameters}
First,  $B$-field parameters common to both the cell-like and Gaussian turbulent morphologies, i.e. the magnetic field strength and $r_\mathrm{max}$, are scrutinised. 
We leave $n_0^\mathrm{ICM}$, $\beta_\mathrm{ICM}$, and $\eta_\mathrm{ICM}$ fixed to their fiducial values. 
In the ALP mass range we are considering here, varying the electron density within a reasonable range
does not have an effect on the conversion probability, cf. eq. \eqref{eqn:ecrit}.
The latter two parameters effectively increase or decrease the magnetic field strength over the cluster,
and thus we restrict ourselves to vary the overall $B$-field normalisation. 

\begin{figure}
\centering
\includegraphics[width = .95 \linewidth]{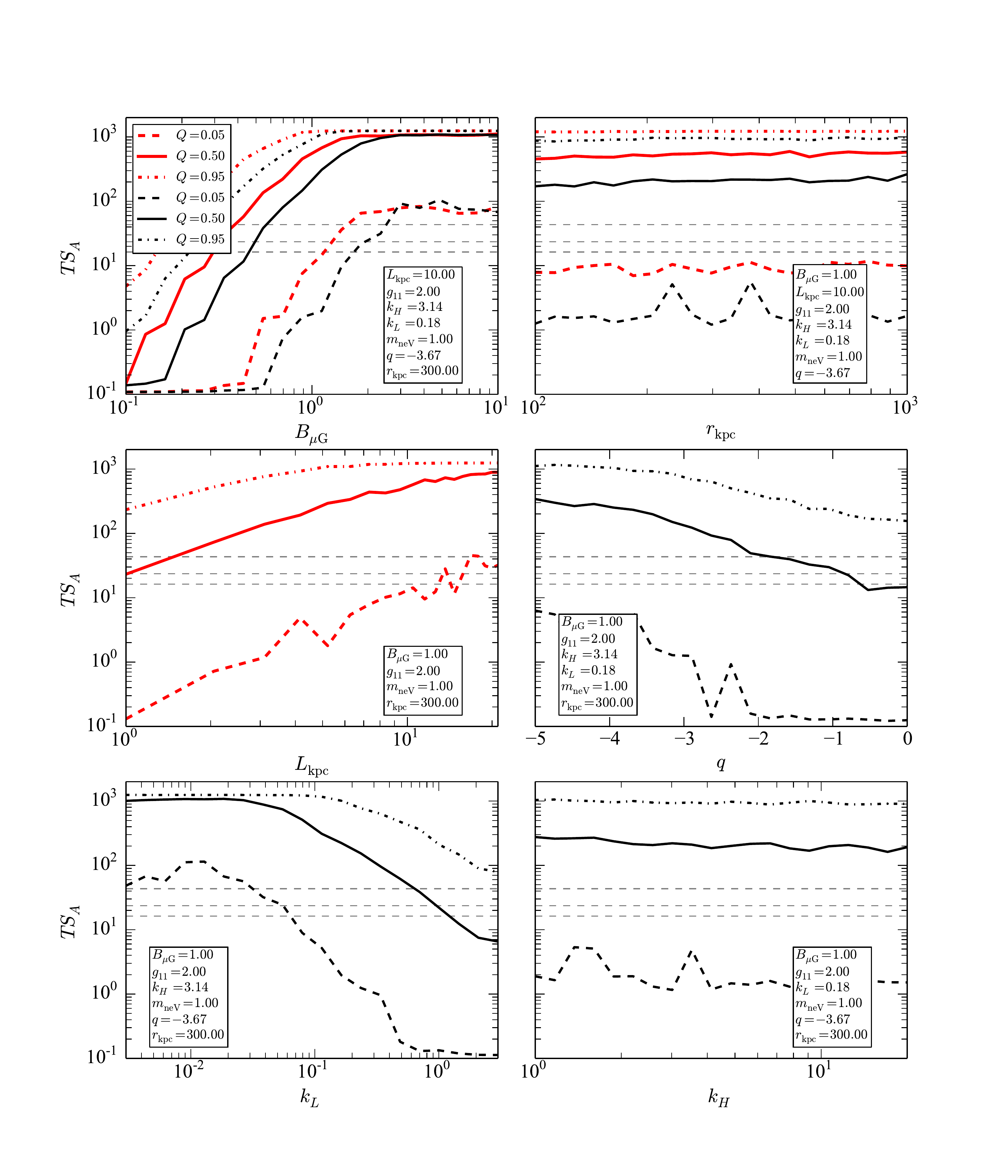}
\caption{Dependence of the sensitivity to ALP induced boosts on ICM parameters. 
The different line styles show the different quantiles of the random magnetic fields. The results of the Gaussian turbulent field are shown 
in black, while for the cell-like morphology results are shown in red. The grey dashed lines give the $\mathit{TS}_A$ values 
that correspond to the 2, 3, and 5\,$\sigma$ confidence levels.
The fluctuation for $Q = 0.05$ visible in all panels is due to the limited statistics of the 1000 random $B$-field simulations. 
See text for further details.
}
\label{fig:b+r}
\end{figure}

In the top panels of figure \ref{fig:b+r} we show the dependence of $\mathit{TS}_A$ on $B_0^\mathrm{ICM}$ and $r_\mathrm{max}$ for 
the different quantiles of the random magnetic fields. 
The $B$-field strength is varied between 0.1 and 10\,$\mu$G.
Such low values for the magnetic field have been inferred from Faraday rotation measure observations
 of NGC\,0315 situated in a low density galaxy group \cite{laing2006}
while the high central magnetic field values have been determined 
in massive cool-core clusters, such as the Hydra A or the Persues cluster \cite{taylor2006,kuchar2011}.
For both morphologies (black lines: Gaussian turbulent; red lines: cell-like $B$ field) 
and all quantiles, the same behaviour is observed:
Below a certain field strength, the spectral differences are to minute to be distinguishable from the no-ALP case. 
Increasing the magnetic field leads to a sharp rise in  $\mathit{TS}_A$ which saturates when on average $1/3$ of the $\gamma$ rays 
have converted into ALPs in the cluster. The amount of ALPs reconverting into photons is determined by the Galactic magnetic field model only. 
Even for pessimistic $B$-field realisations (the $Q = 0.05$ quantile), with our suggested method, the ALP induced effect can be detected at the $3\,\sigma$ and $5\,\sigma$ level for $B_0^\mathrm{ICM} \sim 2\,\mu$G for the cell-like and turbulent morphologies, respectively
(the $\mathit{TS}_A$ values that correspond to  2, 3, and $5\,\sigma$ confidence levels are indicated as grey dashed lines and are derived
from a $\chi^2_\nu$ distribution with $\nu = 6$, see Section \ref{sec:stats}).

Maximum cluster radii between 100\,kpc and 1\,Mpc are tested. 
With all other parameters set to their fiducial values, the experimental sensitivity is almost independent 
of the radius. 
In the cell-like model, this is expected: For $r_\mathrm{max} = 100$\,kpc the condition in eq. \eqref{eqn:effic-mix}
remains fulfilled and on average  $\sim25$\,\% of the photons convert into ALPs.
This is a promising result: also AGN in small scale clusters can be used to search for ALPs. 
Furthermore, this shows that turbulent fields at 100\,kpc scales in AGN jet structures can lead to a sizeable fraction of ALPs in the beam.

The dependence of the $\mathit{TS}_A$ values on the coherence length (cell-like model) and on the power-law index $q$ of the 
turbulence spectrum (Gaussian turbulence model) are shown in the central panels of figure \ref{fig:b+r}.
More turbulent fields, i.e. smaller coherence lengths and smaller values of $q$, lead to smaller effects of ALPs on {\gr} spectra. 
For the cell-like magnetic field, this can again be understood with the help of eq. \eqref{eqn:effic-mix}.
The coherence length $L_\mathrm{coh}$ in the cell-like case can be compared to the correlation length $\Lambda_c$ in the turbulent case \cite[e.g.][]{murgia2004}
 \begin{equation}
\Lambda_c = \frac{1}{C(0)}\int \limits_0^\infty d x_3\, C(x_3) ,
 \end{equation}
 with $C(x_3)$ given by eq. \eqref{eq:correlLOS}. For the fiducial parameters, one finds $\Lambda_c \sim 3.05\,$kpc which decreases
 to $\sim0.57$\,kpc for $q=0$ (white noise). 
 The general trend that the turbulent $B$ field tends to have a smaller impact on $\gamma$-ray spectra than the cell-like morphology
 can also be explained in this picture since $\Lambda_c < L_\mathrm{coh}$ for the fiducial set up\footnote{
 One should be cautious, however, with this interpretation. 
 For $q < -3$, $\Lambda_c$ tends to $1 / k_L$, so that large-scale fluctuations dominate, whereas for $q > -2$, $\Lambda_c \to 1 / k_H$. 
 Only in the latter case when large-scale correlations are absent can $\Lambda_c$ be confused with the domain size $L_\mathrm{coh}$. 
 In general the two morphologies will lead to different results.
 }.
 
Moreover, we investigated the behaviour of the $\mathit{TS}_A$ values with changing IR and UV cut off, $k_L$ and $k_H$
(bottom panels of figure \ref{fig:b+r}). 
Increasing $k_L$ has a very similar effect as increasing $q$: The field becomes more turbulent 
as the maximum turbulence scale $\Lambda_\mathrm{max} = 2\pi / k_L$ decreases, leading to smaller values of $\Lambda_c$.
A degeneracy between $k_L$ and $q$ is also frequently encountered when turbulent $B$-field models are fitted to observations \cite{feretti2012}.
On the other hand, lower values of $k_L$ induce slowly-varying modes in the magnetic field.
As the $B$ field is modulated by the electron density (cf. eqs. \eqref{eqn:bicm} and \eqref{eqn:nicm}) those modes are damped for $r > r_\mathrm{core}$ resulting in a saturation of $\mathit{TS}_A$ at $k_L \lesssim 2\pi / r_\mathrm{core}$.
The $\mathit{TS}_A$ values remain almost constant for varying $k_H$, since scales longer than the inverse of the oscillation length do not contribute 
to the turbulence and we have chosen $q < -3$, i.e. large scale turbulences dominate.
  
\subsection{Jet magnetic-field parameters}
 
 \begin{figure}
\centering
\includegraphics[width = .9 \linewidth]{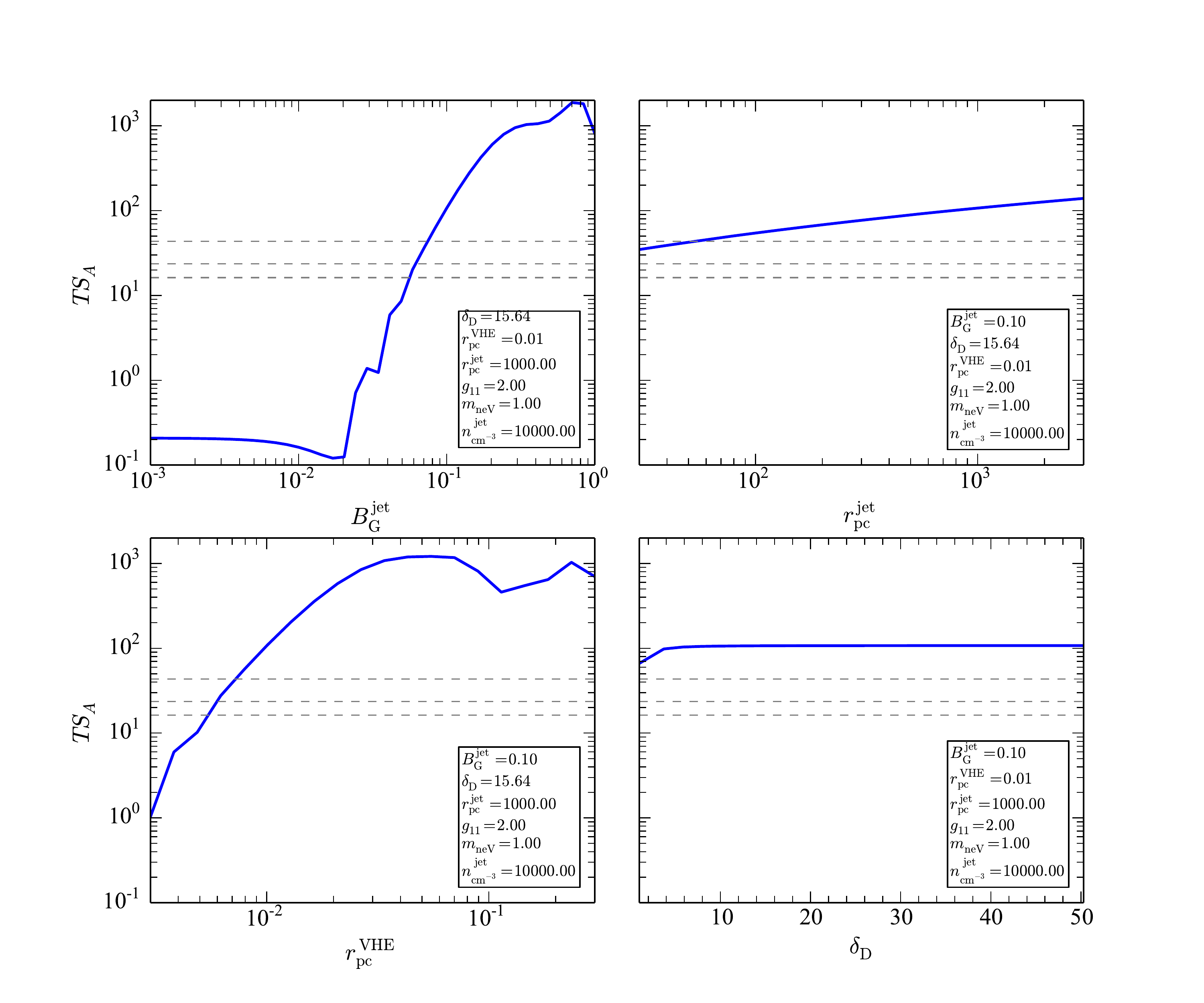}
\caption{Dependence of $\mathit{TS}_A$ on different parameters of the coherent jet magnetic field of a BL Lac-type object.
}
\label{fig:jet}
\end{figure}

In the jet magnetic-field scenario, the choices for the $B$-field normalisation, maximum jet length, 
distance of the VHE emission site to the central engine, and the Doppler factor 
are tested. The power-law index for $B^\mathrm{jet}$, $\eta_\mathrm{jet}$, will be kept constant at its fiducial value. 
The photon-ALP mixing at energies for which $\tau > 2$ ($E \gtrsim 423$\,GeV) will not be affected by the plasma effect, as long as the electron density $n_0^\mathrm{jet} \lesssim 2.5\times10^9\,\mathrm{cm}^{-3}$, see eq. \eqref{eqn:ecrit}. 
This condition is easily fulfilled with electron densities for $r > r_\mathrm{VHE}$ reported in the literature \cite[e.g.][]{lobanov1998,osullivan2009}.
As in the ICM case, a strong dependence of the $\mathit{TS}_A$ values on the magnetic-field strength is observed
 (see the top-left panel of figure \ref{fig:jet}).
 A detection with high confidence is only feasible for $B_0^\mathrm{jet} \gtrsim 0.07$\,G.
 For field strength close to 1\,G, the QED effect becomes important 
 and the maximum energy (cf. eq. \eqref{eqn:emax}) of the SMR falls below 100\,GeV at $r = r_\mathrm{VHE}$ in 
 the lab frame at $\sim0.7$\,G.
 Consequently, the conversion probability starts to oscillate which causes the sudden spikes in the $\mathit{TS}_A$
 values at $B^\mathrm{jet}_0 \gtrsim 0.2\,$G. 
 A mild dependence on the maximum distance scale of the coherent jet $B$ field is evident from the top-right panel of figure \ref{fig:jet}.
 Interestingly, even $B$ fields coherent up to 100\,pc only, will result in a significant detection of the {\gr} boost.
 In contrast, the position of the VHE emission site has a strong effect on the $\mathit{TS}_A$ values (bottom-left panel of figure \ref{fig:jet}).
For the fiducial parameters chosen here, the $\mathit{TS}_A$ values are almost independent of the Doppler factor (bottom-right panel of figure \ref{fig:jet}). 

\subsection{ALP parameters}
\begin{figure}
\centering
\includegraphics[width = .9 \linewidth]{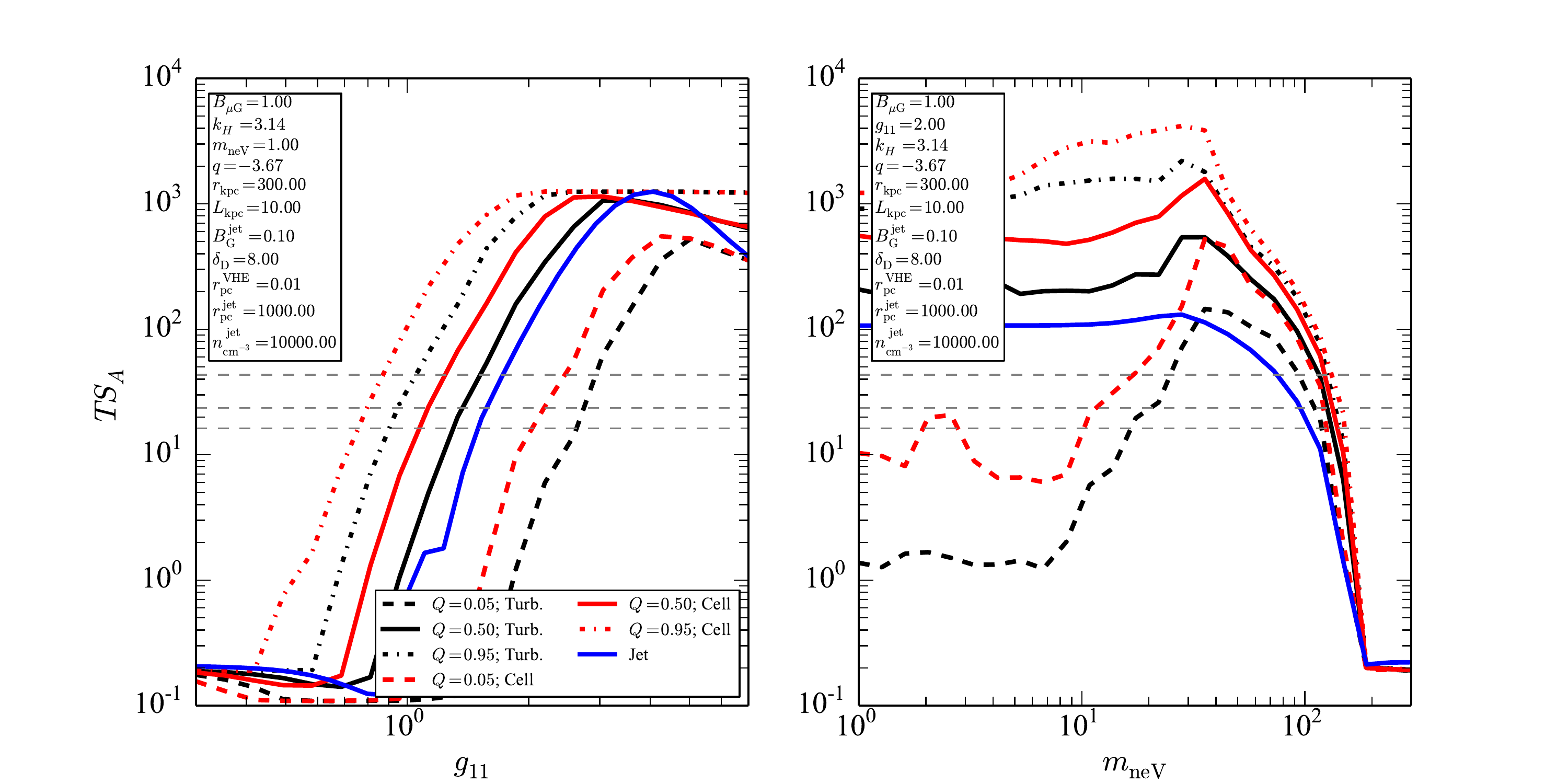}
\caption{Dependence of $\mathit{TS}_A$ on the photon-ALP coupling (\textit{left}) and ALP mass (\textit{right}).}
\label{fig:g+m}
\end{figure}

Eventually, we examine the influence of the photon-ALP coupling and the ALP mass on the sensitivity estimates (figure \ref{fig:g+m}). 
The photon-ALP mixing depends on the product of $B$ and $g_{a\gamma}$ through $\Delta_{a\gamma}$ 
and this complete degeneracy explains the identical behaviour of $\mathit{TS}_A$ with varying $g_{a\gamma}$ (left panel of figure \ref{fig:g+m}) and $B$ (top-left panel of figure \ref{fig:b+r} and top-left panel of figure \ref{fig:jet}). 
The $\mathit{TS}_A$ values rise by $\sim4$ orders of magnitude for a one order of magnitude increase in $g_{a\gamma}$ or $B$.
Again, a saturation at high $g_{a\gamma}$ values is observed. 
For small couplings, the sensitivity of the CTA-like array is not sufficient to observe a reduced opacity in the mock AGN spectrum.

An interesting behaviour is observed if the ALP mass is increased (right panel of figure \ref{fig:g+m}). 
Naively, one would expect that the $\mathit{TS}_A$ values stay constant until $m_a$ becomes large enough so that $E_\mathrm{crit} > 423\,$GeV.
Further increasing the mass should lead to a declining $\mathit{TS}_A$ value as 
more and more data points fall outside the SMR. 
This general trend is indeed observed for $m_\mathrm{neV}\gtrsim 40$.
For masses between $\sim10$ and $\sim40$\,neV, the $\mathit{TS}_A$ values strongly increase.
The reason for this is the oscillatory behaviour of $P_{\gamma\gamma}$ in the transition to the SMR. 
Thanks to the fully energy dependent treatment of the photon-ALP oscillations adopted here, 
the likelihood ratio test is sensitive to these spectral features.

\section{Summary and Conclusion}
\label{sec:summary}
In this article, we have discussed oscillations of photons into axion-like particles in turbulent and coherent magnetic fields and its impact on the detection of $\gamma$ rays in the optical thick regime. 
We have derived specific formulas to calculate the  components of a magnetic field with gaussian turbulence. 
The sensitivity for a CTA-like array to detect a boost in the photon flux 
has been calculated for 
wide variety of
magnetic-field and ALP parameters
using the Asimov data set. 
While the Asimov data set gives a good estimator for the median of many Monte-Carlo simulations,
fluctuations of real data can of course be larger \cite{cowan2011}.

With the suggested method, modifications of the spectra should be detectable for couplings $g_{11} \gtrsim 2$ and ALP masses
$m_\mathrm{neV} \lesssim 100$ for a viable range of the ambient magnetic field strength  given a 20\,hours observation of a flaring AGN 
with properties similar to the object PG\,1553+113 with an intrinsic spectrum that follows a power-law extrapolation up to $\sim 7.4$\,TeV
(corresponding to $\tau = 12$).
These ALP parameters are also well in range of the future laboratory experiment ALPS II \cite{alpsII} and the next  generation Helioscope, IAXO \cite{irastorza2013}.

As expected, the test statistic strongly depends on the magnetic-field strength, the photon-ALP coupling, as well as on the ALP mass,
regardless of the magnetic-field scenario.
Only a mild dependence is found on the size of the $B$-field region within the tested values, for both the BL Lac jet and the galaxy cluster $B$ fields. 
In accordance with ref. \cite{tavecchio2014}, we find a strong dependence of the ALP effect on 
the distance of the VHE emission zone to the central engine when mixing in a BL Lac jet is considered. 
If the jet magnetic field is below $\sim 0.05$\,G or the emitting site is close to the central 
black hole, $r_\mathrm{VHE} \lesssim 0.01$\,pc,  a spectral modification will become difficult to detect
with the assumed observation. 
The same is true if the source is located in a low-density type environment as 
observed around the radio galaxy NGC\,0315 where the 
intra-cluster magnetic fields can be as low as $\sim0.1\,\mu\mathrm{G}$ \cite{laing2006}.

We have derived formulas to compute the transversal component of a magnetic field with gaussian turbulence. 
Compared to a simple cell-like models, very similar results are obtained for the sensitivity.
The turbulent magnetic fields show a strong dependence on the assumed coherence length, the power-law index, and assumed minimum 
and maximum scales of the turbulence spectrum.

Extension to higher ALP masses are possible with observations at higher energies above tens of TeV, e.g. of close-by AGN.
This energy range is also probed with the High Altitude Water Cherenkov Experiment (HAWC, \cite{hawc2013})
 which can detect $\gamma$ rays up to $\sim 100$\,TeV. 
An enhanced {\gr} flux beyond these energies could be detected with the  planned HiSCORE experiment \cite{tluczykont2011}. 

 The suggested likelihood test is not limited to the search for ALP signatures but
 is applicable to any process that modifies the optical depth predicted by standard EBL absorption models.
 For example, cosmic rays accelerated in AGN could effectively transport energy 
 close to Earth and then initiate an electromagnetic cascade \cite{essey2010,essey2010b,essey2011}. 
 The generated secondary {\gr} flux could also explain hints for a spectral hardening in VHE spectra \cite{essey2012}.

For future work, the full IRFs for CTA including the energy dispersion and the PSF should be used, together
with a systematic scan over a grid of the ALP mass and coupling values. 
With a full set of IRFs one should also investigate the dependence of the sensitivity to different 
models of the EBL and the Galactic magnetic field, the energy dispersion, and a possible curvature in the intrinsic spectrum.
Moreover, the likelihood in eq. \eqref{eqn:likelihood} can easily be extended to include several AGN.
Ultimately, the observations of many sources in the optical thick regime at different energies and redshifts
 are necessary to exclude source intrinsic effects.
 
\begin{appendix}
\section{Derivation of transverse components of turbulent magnetic fields}
\label{app:bfield}

In this appendix we show how to simulate a realisation of the transversal component along a line of sight for an isotropic and homogeneous gaussian turbulent magnetic field with zero mean and root-mean-square (r.m.s.) value ${\cal B}^2$. For such a field each component can be expanded in Fourier modes:
\begin{equation}
B_i({\mathbf x})=\int\frac{d^3 k}{(2\pi)^3}\,
{\tilde B}_i({\mathbf k})e^{i({\mathbf k}\cdot{\mathbf x}+\psi_i({\mathbf k}))}\, ,
\end{equation}
where the ${\tilde B}_i({\mathbf k})$ are real even functions [${\tilde B}_i(-{\mathbf k})={\tilde B}_i({\mathbf k})$]. The correlation function for the Fourier modes for the components of an isotropic and homogeneous magnetic field can be written as:
\begin{equation}
\langle{\tilde B}_i({\mathbf k}){\tilde B}_j({\mathbf k}')\rangle=
(2\pi)^6\left[ M(k)P_{ij}({\mathbf k})+E(k)i\epsilon_{ijl}\frac{k_l}{k}\right]\delta^3({\mathbf k}-{\mathbf k}')\, ,
\end{equation}
were $k=|{\mathbf k}|$. The tensor
\begin{equation}
P_{ij}({\mathbf k})=\delta_{ij}-\frac{k_i k_j}{k^2}
\end{equation}
assures the condition $\nabla\cdot{\mathbf B}=0$, and $M(k)$ is the spectrum of ``symmetric'' perturbations normalized in such a way that the r.m.s.\ coincides with ${\cal B}^2$:
\begin{equation}
{\cal B}^2=\langle B_i({\mathbf x})B^i({\mathbf x})\rangle=
8\pi \int_0^\infty dk\, k^2 M(k)\, .
\label{eq:norm}
\end{equation}
$M(k)$ is thus the energy per unit volume of the phase space $d^3x\,d^3k$. 
In the following, we neglect the ``helical'' component $E(k)$ since it does not give a contribution to the quantities that we want calculate. 
If we choose the $x_3$ axis as the line of sight, we have for each of the transversal components (we choose $B_1$ without loss of generality)
\begin{eqnarray}
\langle B_1({\mathbf x})B_1({\mathbf x}+x_3{\hat{\mathbf e}}_3)\rangle&=&
\int d^3 k\, M(k)\left(1-\frac{k_1^2}{k^2}\right)e^{ik_3 x_3}\nonumber \\
&\equiv&\int \frac{dk_3}{2\pi}\,{\tilde\varepsilon}_\perp(k_3)e^{ik_3 x_3}\, ,
\end{eqnarray}
where ${\tilde\varepsilon}_\perp(k_3)$ is the correlation function on the line of sight for the transversal $B$ components, that can be expressed as 
\begin{equation}
{\tilde\varepsilon}_\perp(k_3)=2\pi\int dk_1dk_2\,  M(k)\left(1-\frac{k_1^2}{k^2}\right)\, ,
\end{equation}
with $k^2=k_1^2+k_2^2+k_3^2$. In cylindrical coordinates ($k_1=k_\perp\cos\phi$, $k_2=k_\perp\sin\phi$), the integral becomes
\begin{equation}
{\tilde\varepsilon}_\perp(k_3)=2\pi\int_0^\infty dk_\perp\int_0^{2\pi}d\phi\,  k_\perp M(k)\left(1-\frac{k_\perp^2\cos^2\phi}{k^2}\right)\, .
\end{equation}
Integrating over $\phi$ and with the change of variable $k^2=k_\perp^2+k_3^2$ we obtain:\footnote
{
In the same way, the longitudinal component can be calculated as follows
\begin{equation}
{\tilde\varepsilon}_{||}(k_3)=4\pi^2\int_{|k_3|}^\infty dk\,  k M(k)\left(1-\frac{k_3^2}{k^2}\right)\, .
\end{equation}
}
\begin{equation}
{\tilde\varepsilon}_\perp(k_3)=2\pi^2\int_{|k_3|}^\infty dk\,  k M(k)\left(1+\frac{k_3^2}{k^2}\right)\, .\label{eqn:trans-corr}
\end{equation}
The spatial correlation function along the line of sight for any of the transversal components of the magnetic field is given by
\begin{equation}
C(x_3)=\langle B_\perp({\mathbf x})B_\perp({\mathbf x}+x_3{\hat{\mathbf e}}_3)\rangle=
\int_0^\infty\frac{dk}{\pi}{\tilde\varepsilon}_\perp(k)\cos k x_3\, ,
\label{eq:correlLOS}\end{equation}
where $B_\perp\equiv B_{1,2}$ and we have used the fact that ${\tilde\varepsilon}_\perp(k)$ is a real even function.

Following \cite{majda1999}, the transversal components $B_{1,2}(x_3)$ along a given line-of-sight can be simulated by the discrete Fourier expansion (where the modes are not necessarily equally spatiated):
\begin{equation}
B_\perp(x_3)\simeq\sum_{n=1}^{N_k}  b_n\left[\xi_n\cos k_n x_3+\eta_n\sin k_n x_3\right]\, .
\label{eq:spectrumlos}
\end{equation}
where $\xi_n$, $\eta_n$ are normally distributed independent variables with $0$ mean and unit variance ($\langle \xi_{n_1}\xi_{n_2}\rangle=\langle \eta_{n_1}\eta_{n_2}\rangle=\delta_{n_1 n_2}$,  $\langle \xi_{n_1}\eta_{n_2}\rangle=0$). The correlation function $C(x_3-x_3')=\langle B_\perp(x_3)B_\perp(x_3')\rangle$ is thus given by
\begin{equation}
C(x_3-x_3')\simeq\sum_{n=1}^{N_k} b_n^2 \cos \left(k_n (x_3-x_3')\right)\, .
\end{equation}
From the discretized eq.~(\ref{eq:correlLOS}) we obtain
\begin{equation}
b_n=\sqrt{\frac{{\tilde\varepsilon}_\perp(k_n)\Delta k_n}{\pi}}\, ,
\end{equation}
where $\Delta k_n$ is the width of the interval containing $k_n$. Finally, using the Box-Muller theorem \cite{box1958}, we can rewrite eq.~(\ref{eq:spectrumlos}) as eq. \eqref{eqn:bfield}.

\begin{figure}[t!b]
\centering
\includegraphics[width = .7 \linewidth]{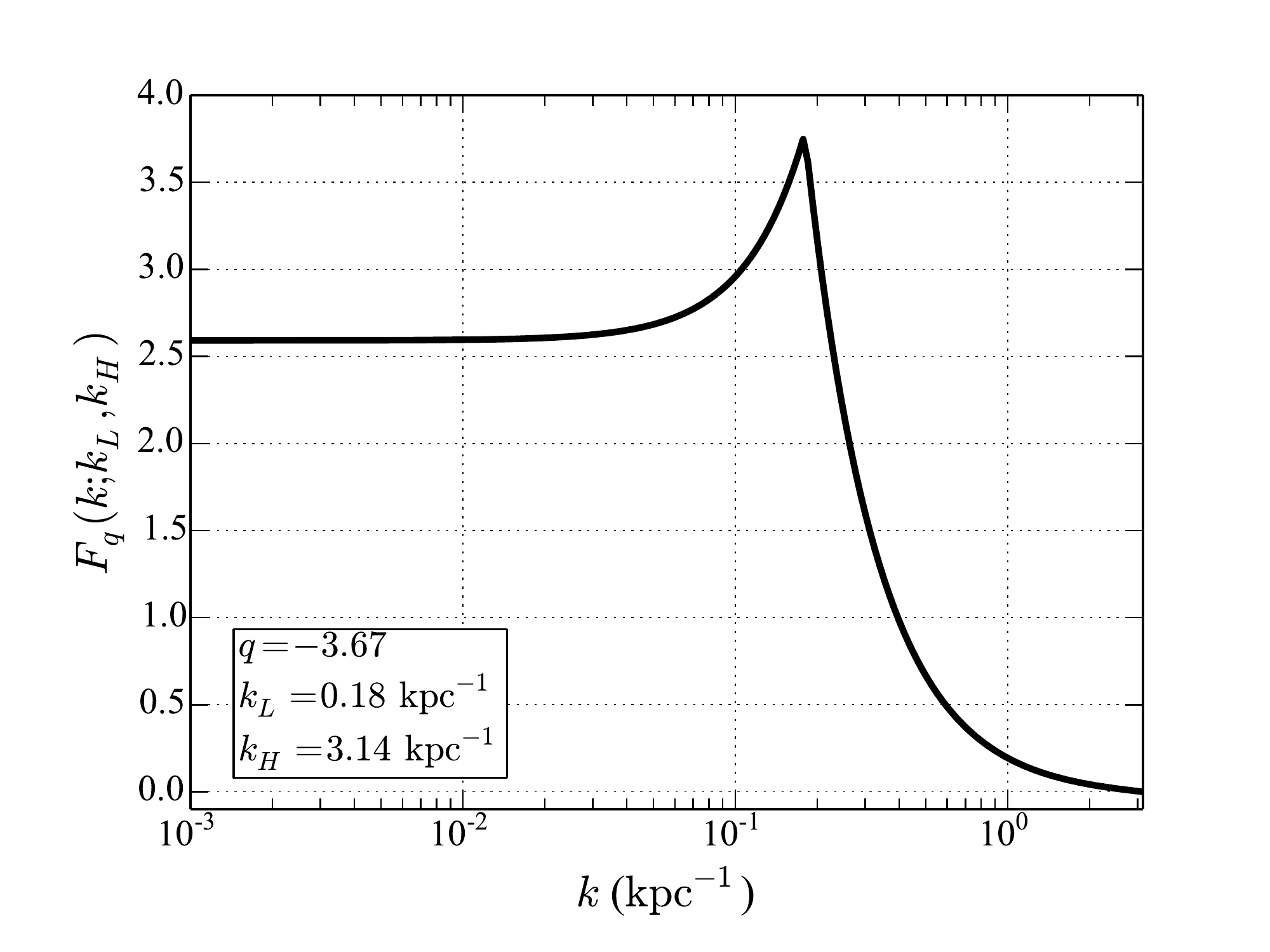}
\caption{The function $F_q(k; k_L,k_H)$ for the fiducial model parameters chosen in the galaxy cluster mixing scenario.}
\label{fig:F_x}
\end{figure}

Next, we consider the case of a power-law spectrum with an ultraviolet (UV) cut-off $k_H$ and infrared (IR) cut-off $k_L$: 
\begin{equation}
M(k) = A k^q,\quad\mathrm{for}\, k_L \leqslant k \leqslant k_H,\, 0\,\mathrm{otherwise}.
\end{equation}
The normalization condition~(\ref{eq:norm}) gives
\begin{equation}
A = \frac{\mathcal{B}^2}{8\pi}
\begin{cases}
\frac{q +3}{k_H^{q + 3} - k_L^{q + 3}} & q \neq -3, \\
\ln\frac{k_L}{k_H} & q = -3
\end{cases}.
\end{equation}
After a straightforward calculation, from eq.~(\ref{eq:spectrumlos}) we obtain eq. \eqref{eqn:eps-trans} with
\begin{equation}
F_q(k,k_L,k_H)= \frac{f_{q+2}(k',k_H) + k^2 f_{q}(k',k_H)}{f_{q+3}(k_L,k_H)}\label{eqn:Fx}
\end{equation}
for $k<k_H$ and $F_q(k,k_L,k_H)=0$ for $k\geq k_H$, with $k'={\rm max}(k,k_L)$ and
\begin{equation}
f_q(s_1,s_2) =\left\{
\begin{array}{ll}
(s_2^q-s_1^q)/q, & \mathrm{if}\,q\neq 0, \\
\ln\left(s_2/s_1\right), & \mathrm{if}\,q = 0.
\end{array}\right.
\end{equation}
The case $q = 0$ corresponds to white noise. 
For $k < k_L$,  $F_q(k; k_L, k_H)$ tends to a constant, as can be seen from splitting up the integral in eq. \eqref{eqn:trans-corr}. 
The function $F_q(k; k_L, k_H)$ is shown in figure \ref{fig:F_x} for the fiducial values of a galaxy cluster adopted here (cf. Table \ref{tab:fields}). 
One random $B$-field realisation is shown in figure \ref{fig:fields}.
Notice that the tensor $P_{ij}$ is responsible for the ``blue'' component $k^2$ always present in the line of sight spectrum.

\begin{figure}[t!b]
\centering
\includegraphics[width = .7 \linewidth]{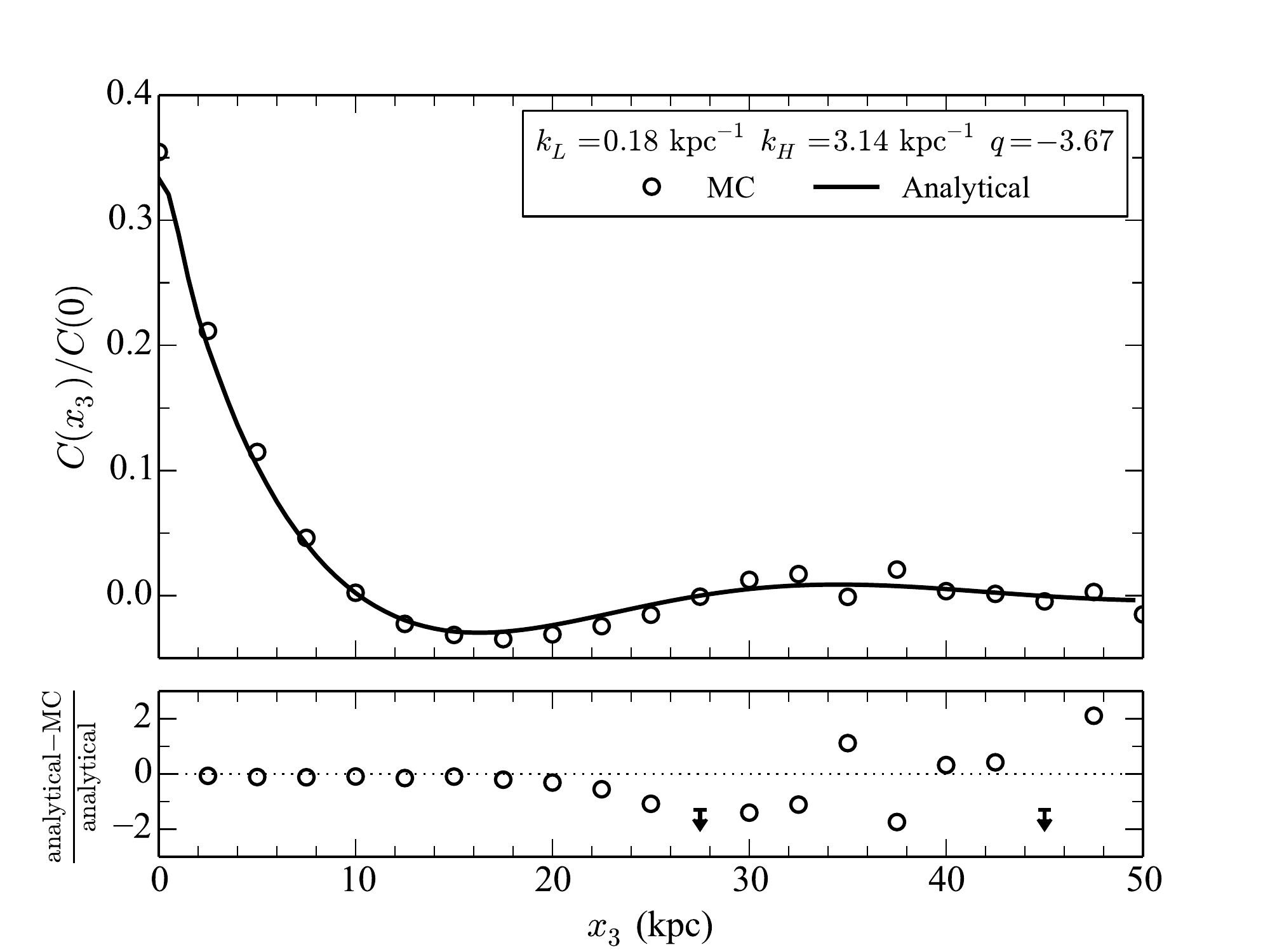}
\caption{Comparison between the analytical correlation function and Monte-Carlo simulations. Shown are the calculations with
 $N_k = 84$ and $k \in [10^{-3}k_L,k_H]$. 
An equidistant logarithmic spacing is chosen.
The lower panel displays the residuals defined as the difference between
the analytical calculations and the Monte-Carlo simulations divided by the analytical result. 
For two points, the residuals are
 $\sim-10$, indicated by the arrows in the plot.}
\label{fig:corr}
\end{figure}

In figure~\ref{fig:corr} we show the spatial correlation function as function of $x_3$ for a ``Kolmogorov-like''
spectrum with index $q=-11/3$ normalized with the zero distance correlation (that is the r.m.s. of the magnetic field). 
We also show the results obtained with a Monte-Carlo generation of the turbulent magnetic field.
A good agreement is found between the analytical prescription and the simulations. 

\end{appendix}

\acknowledgments
We would like to thank Christian Farnier, Fabrizio Tavecchio, Marco Roncadelli, and Giorgio Galanti for providing valuable comments to the manuscript.
D.M. acknowledges support from the Istituto Nazionale di Fisica Nucleare (INFN,
Italy) through the ``Astroparticle Physics'' (TAsP) research project.
J.C. is a Wallenberg Academy Fellow.

\bibliographystyle{JHEP}
\bibliography{meyer_ALP_transparency,vhe_spectra,galaxy_clusters}
\end{document}